\documentclass{aa}  
\usepackage{graphicx}
\usepackage{txfonts}
\usepackage{color}
\usepackage{float}
\usepackage{courier}
\usepackage{tabularx}
\newcolumntype{L}[1]{>{\hsize=#1\hsize\raggedright\arraybackslash}X}%
\newcolumntype{R}[1]{>{\hsize=#1\hsize\raggedleft\arraybackslash}X}%
\newcolumntype{C}[2]{>{\hsize=#1\hsize\columncolor{#2}\centering\arraybackslash}X}%
\begin{document}

   \title{The mean H$\alpha$ EW and Lyman-continuum photon production efficiency for faint $z\approx4-5$ galaxies}


   \author{Daniel Lam
          \inst{1}
          \and
          Rychard J. Bouwens\inst{1}
          \and
          Ivo Labb\'{e}\inst{2}
          \and
          Joop Schaye\inst{1}
          \and
          Kasper B. Schmidt\inst{3}
          \and
          Michael V. Maseda\inst{1}
          \and
          Roland Bacon\inst{4}
          \and
          Leindert A. Boogaard\inst{1}
          \and
          Themiya Nanayakkara\inst{1}
          \and
          Johan Richard\inst{4}
          \and
          Guillaume Mahler\inst{5}
          \and
          Tanya Urrutia\inst{3}
          }

   \institute{Leiden Observatory, Leiden University, NL-2300 RA Leiden, Netherlands\\
              \email{daniellam@strw.leidenuniv.nl}
              \and Centre for Astrophysics and SuperComputing, Swinburne, University of Technology, Hawthorn, Victoria, 3122, Australia
              \and Leibniz-Institut für Astrophysik Postdam (AIP), An der Sternwarte 16, D-14482 Potsdam, Germany
              \and Univ Lyon, Univ Lyon1, Ens de Lyon, CNRS, Centre de Recherche Astrophysique de Lyon UMR5574, F-69230, Saint-Genis-Laval, France
              \and Department of Astronomy, University of Michigan, 1085 South University Ave, Ann Arbor, MI 48109, USA
              }

   \date{}

 
\abstract{
We present the first measurements of the Lyman-continuum photon production efficiency $\xi_{\textrm{ion,0}}$ at $z\sim4$-5 for galaxies fainter than 0.2 $L^*$ ($-$19 mag).  $\xi_{\textrm{ion,0}}$ quantifies the production rate of ionizing photons with respect to the UV luminosity density assuming a fiducial escape fraction of zero.   Extending previous measurements of $\xi_{\textrm{ion,0}}$ to the faint population is important, as ultra-faint galaxies are expected to contribute the bulk of the ionizing emissivity.   We probe $\xi_{\textrm{ion,0}}$ to such faint magnitudes by taking advantage of 200-hour depth Spitzer/IRAC observations from the GREATS program and $\approx$300 3<$z$<6 galaxies with spectroscopic redshifts from the MUSE GTO Deep + Wide programs. Stacked IRAC [3.6]$-$[4.5] colors are derived and used to infer the H$\alpha$ rest-frame equivalent widths, which range from 403\r{A} to 2818\r{A}. The derived $\xi_{\textrm{ion,0}}$ is $\log_{10}(\xi_{\textrm{ion,0}} / \textrm{Hz erg}^{-1}) = 25.36 \pm 0.08$ over $-$20.5 < M$_{\textrm{UV}}$ < $-$17.5, similar to those derived for brighter galaxy samples at the same redshift and therefore suggesting that $\xi_{\textrm{ion}}$ shows no strong dependence on $M_{UV}$.  The $\xi_{\textrm{ion,0}}$ values found in our sample imply that the Lyman-continuum escape fraction for $M_{\textrm{UV}} \approx -19$ star-forming galaxies cannot exceed $\approx$8-20\% in the reionization era. 
}

   \titlerunning{IRAC-inferred $\xi_{\textrm{\textrm{ion}}}$}

   \maketitle
%

\section{Introduction}

The reionization of the universe has received significant attention over the last decade. 
Fundamental unanswered questions remain about both the basic time scale of cosmic reionization and the sources which drive the process. 
The most obvious sources to power cosmic reionization are star-forming galaxies \citep[e.g.,][]{bouwens15, robertson15} and quasars/active galactic nuclei \citep[AGN, ][]{madau15}. 
One of the reasons why quantifying their respective contribution to cosmic reionization has been challenging is that we cannot detect ionizing photons from these sources directly.

As such, our best estimates for the ionizing emissivity from galaxies or quasars have been based on the emissivity of these sources in the non-ionizing UV-continuum. 
It has been conventional to convert the galaxy UV luminosity density to an ionizing emissivity (or the rate of ionizing photons per unit volume that reaches the intergalactic medium) $\dot{n}_{\textrm{ion}}$:
\begin{equation}\label{eq:definition}
\dot{n}_{\textrm{ion}} = \frac{\rho_{\textrm{UV}}}{f_{\textrm{esc,UV}}} \hspace{0.1cm} \xi_{\textrm{ion}} \hspace{0.1cm} f_{\textrm{esc,LyC}} \hspace{0.1cm} \mathrm{, }
\end{equation}
where $\rho_{\textrm{UV}}$ is the total UV (1500\r{A}) luminosity density, $\xi_{\textrm{ion}}$ the ionizing photon production efficiency per unit UV luminosity, and $f_{\textrm{esc,UV}}$ and $f_{\textrm{esc,LyC}}$ the fraction of light that is able to escape the galaxy unabsorbed in the non-ionizing UV and ionizing wavelengths, respectively.

Traditionally, the Lyman-continuum photon production efficiency $\xi_{\textrm{\textrm{ion}}}$ has been estimated by extrapolating from either the rest-frame UV-continuum slope $\beta$ \citep{robertson13,duncan15,bouwens16b} or from synthetic stellar population models \citep{madau99}. 
In \citet{bouwens16} (hereafter B16), however, it was shown that $\xi_{\textrm{ion}}$ can be directly inferred from Spitzer/IRAC-based estimates of the H$\alpha$ fluxes.  As demonstrated by \citet{shim11} and \citet{stark13}, observations with Spitzer/IRAC can be used to infer the H$\alpha$ fluxes as H$\alpha$ falls in the 3.6 $\mu$m (3.8$<$z$<$5.0) and 4.5 $\mu$m bands (5.1$<$z$<$6.6). 
Given that the vast majority of ionizing photons produced by stars in a galaxy ionize neutral hydrogen which cascades down to produce H$\alpha$, the production rate of ionizing photons can be deduced quite straightforwardly from the H$\alpha$ line flux using quantum mechanics \citep{leitherer95}.

The availability of two new data sets make it possible to improve upon the estimates in B16, particularly by allowing us to push fainter.  The first data set is the Multi Unit Spectroscopic Explorer (MUSE) GTO data over the HUDF and GOODS-S fields \citep{bacon17,herenz17}. 
From these data, it has been possible to construct very large samples of $z\sim$3-6 galaxies, all with spectroscopic redshifts \citep{inami17,herenz17}. 
The advantage of using a purely spectroscopic sample, rather than the photometric samples used in some earlier studies \citep{smit16,marmol16,rasappu16}, is that we can derive the H$\alpha$ fluxes, while ensuring that the H$\alpha$ line always falls in the desired IRAC filter.

The second of these data sets is the 3.6 $\mu$m + 4.5 $\mu$m Spitzer/IRAC GREATS observations \citep{labbe14}, which has an integration time of 200 hours in the deepest parts, resulting in photometric depths at least 0.4 mag deeper than used in the earlier studies on which B16 is based. 
By taking advantage of the deeper Spitzer/IRAC data and stacking the large numbers of sources we have from MUSE, we can probe the H$\alpha$ line flux in galaxies which are more representative of the overall population, and have a greater contribution to the overall photon budget, than from the rarest, brightest ones.

\begin{table*}
\centering
\caption[]{Data sets utilized in this study. }
\label{table:data}
\begin{tabularx}{\textwidth}{L{0.6} L{1.2} L{1.2}}
\hline
\noalign{\smallskip}
Data set  &  Description  &  Depth \\
\noalign{\smallskip}
\hline
\noalign{\smallskip}
MUSE-Deep  &  A 3D spectroscopic survey that covers 9.92 arcmin$^{2}$ of the HUDF with an average exposure time of 10 hours \citep{bacon17}. A 1 arcmin$^{2}$ area near the center receives an additional 31 hours of exposure time.  &  3$\sigma$ emission line detection limit for a point source ranges from 1.5 to 3.1 $\times$ 10$^{-19}$ erg s$^{-1}$ cm$^{-1}$. \\
\noalign{\smallskip}
\hline
\noalign{\smallskip}
MUSE-Wide  &  A 3D spectroscopic survey that covers 44 1-arcmin$^{2}$ fields (DR1) each with an exposure time of 1 hour \citep{urrutia18}.  &  Detection limit for emission lines of a point source is 8 $\times$ 10$^{-18}$ erg s$^{-1}$ cm$^{-2}$ \r{A}$^{-1}$.  \\
\noalign{\smallskip}
\hline
\noalign{\smallskip}
GOODS Re-ionization Era wide-Area Treasury from Spitzer  &  Ultradeep (GREATS) imaging survey with the Spitzer Space Telescope over the GOODS-South and North (not utilized in this study) fields \citep{labbe14}. The exposure time over our FOV of analysis ranges from $\approx$70 to $\approx$270 hours.  &  5$\sigma$ limiting AB magnitudes of 26.6 and 26.5 averaged over the FOV of this study in the 3.6 $\mu$m and 4.5 $\mu$m band, respectively.  \\
\noalign{\smallskip}
\hline
\noalign{\smallskip}
Hubble Legacy Fields  &  A combination of HST imaging data taken by 31 programs over the GOODS-S field \citep[HLF, ][]{illingworth16}. Includes all optical ACS/WFC filters and all infrared WFC3/IR filters.  The sum of the exposure time for the full data set considered is 1611 hours.  &  5$\sigma$ limiting AB magnitudes of the nine filters (\S 2.3) range from 25.5 (F140W) to 28.3 (F850LP) over our FOV of analysis.  \\
\noalign{\smallskip}
\hline
\end{tabularx}
\end{table*}

In \S2, the observational data are summarized, along with the sample selection criteria. 
\S3 describes our procedure for performing photometry, as well as parameter inference. 
In \S4, we present new measurements of the Lyman-continuum photon production efficiency $\xi_{\textrm{ion}}$ based on the GREATS data and MUSE spectroscopic redshift samples. 
Finally, we discuss our results in \S5 and provide a brief summary in \S6. 
We assume $\Omega_0 = 0.3$, $\Omega_{\Lambda} = 0.7$, and $H_0 = 70\,\textrm{km/s/Mpc}$ throughout this paper. 
All magnitudes are in the AB system \citep{oke83}.

\section{Observational Data and Sample Selection}

In this section, we provide a description of all significant data sets that we utilize for our analysis. Table \ref{table:data} provides a convenient summary.

\subsection{Spectroscopic Samples}

One of the major contributing factors to our being able to push faint in the present study are the large samples of galaxies with spectroscopic redshifts we have from two large MUSE GTO programs: the MUSE-Deep and MUSE-Wide, spectroscopic surveys.\footnote{In principle, with photometric redshifts, we could also segregate faint sources over the HUDF by redshift, but the larger flux uncertainties combined with the unknown impact of the Ly$\alpha$ emission line make the redshift estimates less certain, making the fainter sources significantly less straight forward to include.} 
MUSE-Deep \citep{bacon17} has 116 hours of total exposure time over an area of 3.15$\times$3.15 arcmin$^{2}$ covering the Hubble eXtreme Deep Field \citep[XDF, ][]{illingworth13}. 
MUSE-Deep consists of two components: a shallower $3\times3$ `mosaic' of nine 1$\times$1 arcmin$^{2}$ fields that cover the entire MUSE-Deep region, and a deeper, single 1$\times$1 arcmin$^{2}$ field named 'UDF10' that overlaps with  deep ALMA observations \citep{walter16}. 
We use the MUSE-Deep spectroscopic redshift catalog published by \cite{inami17}, which sources include optical+NIR HST detection as well as blind searches for emission lines in the spectral cube. 

The MUSE-Wide survey \citep{urrutia18} complements the MUSE-Deep by surveying a much larger area to probe rare sources. 
Here we utilize the first data release of MUSE-Wide which consist of 44 1-hour pointings \citep{urrutia18}.
The data cover an area of $\approx$44 arcmin$^{2}$ distributed over the CANDELS-DEEP WFC3 region \citep{koekemoer11}. 
All spectroscopic redshifts from both MUSE-Deep and MUSE-Wide greater than z=2.9 are inferred from the Lyman-alpha line. 

\begin{figure}
  \centering
  \includegraphics[width=8.8cm]{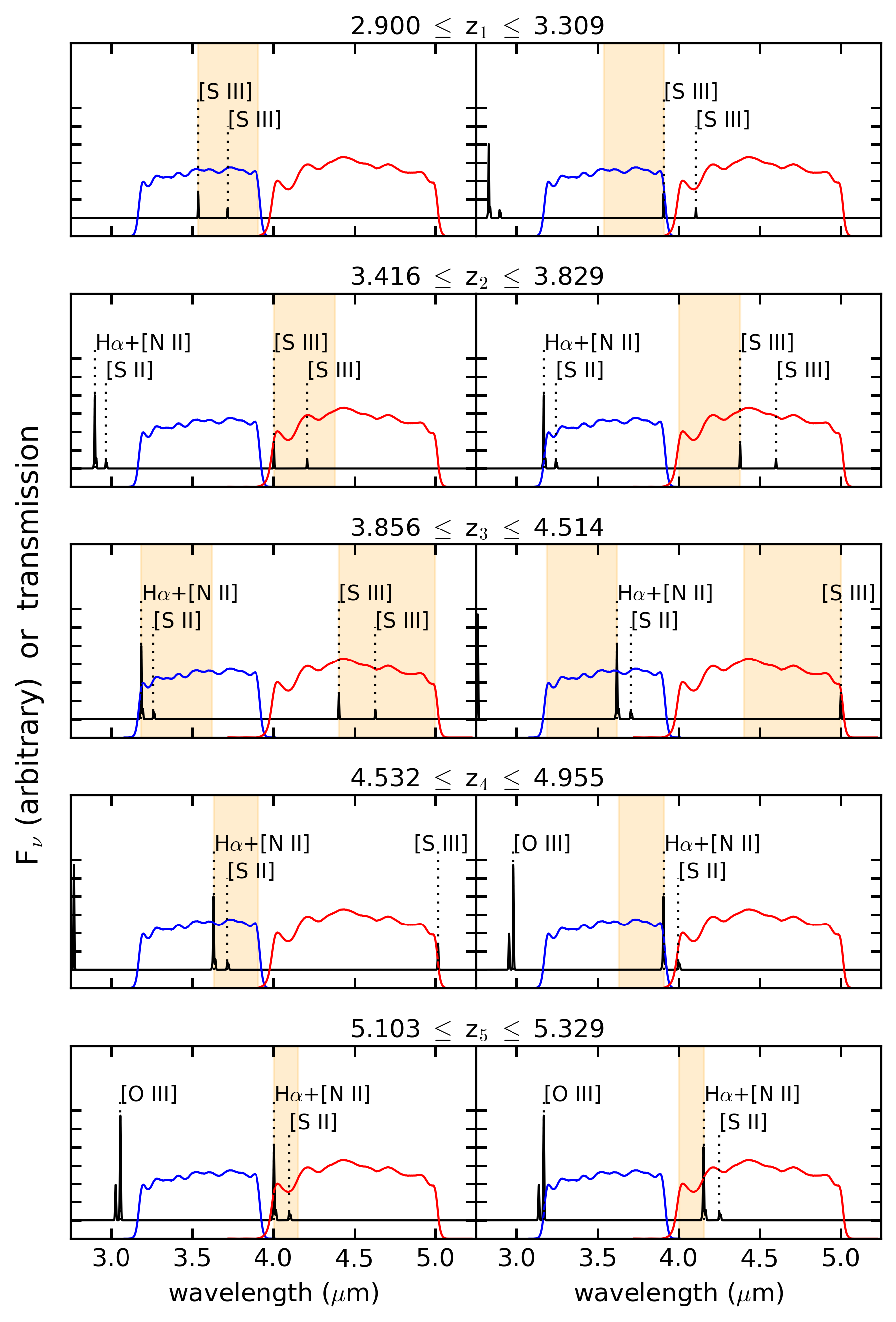}
  \caption{
Illustration of where [S III] 9530.9 \r{A}, H$\alpha$, and [O III] 5006.84 \r{A} lie within the Spitzer/IRAC 3.6 $\mu$m and 4.5 $\mu$m bands at the low and high-redshift ends (\textit{left} and \textit{right} panels, respectively) of the redshift intervals selected for this analysis. 
The transmission curves of the 3.6 $\mu$m and 4.5 $\mu$m bands are shown as the blue and red lines, respectively. 
The orange color-shaded regions denote the wavelengths over which relevant lines (H$\alpha$ and [SIII] 9068.6 \r{A}) lie within our defined redshift windows. 
$z_{1}$ does not extend beyond $z=2.9$ because MUSE sources at $z < 2.9$ are identified from spectral features other than Lyman $\alpha$. }
  \label{fig:halpha}
\end{figure}

To make use of our large sample of spectroscopic redshifts to constrain the flux in the H$\alpha$ line and other lines, we need to segregate sources in redshift windows where a consistent set of strong rest-frame optical emission lines contribute to the IRAC fluxes. 
The most obvious choices of redshift windows are those where the H$\alpha$ line lies in either the 3.6$\mu$m or the 4.5$\mu$m band while no other strong lines are present in the other. 
This criteria is met for sources at $4.532 \leq z \leq 4.955$ (denoted as z$_{4}$) and at $5.103 \leq z \leq 5.329$ (z$_{5}$), as shown in Figure \ref{fig:halpha}. 
We define an emission line to be outside of a filter if its central wavelength lies at half of the `minimum in-band transmission' and inside if it lies at the `minimum in-band transmission'. 
The `minimum in-band transmission' of the IRAC 3.6$\mu$m and 4.5$\mu$m filters are 56.3\% and 54.0\% respectively. 
Beyond z = 5.329, the strong emission line [O III] 5006.84 $\mbox{\AA}$ enters the 3.6 $\mu$m band, while H$\alpha$ is still present in the 4.5$\mu$m band.  
Given the potentially large uncertainties that may result in correcting the 3.6$\mu$m flux for the [OIII] contribution, we do not consider the 141 sources beyond $z = 5.329$. 

The numbers of sources in redshift intervals z$_{4}$ and z$_{5}$ are small due to their narrow widths. 
This can be mitigated by extending the same methodology to lower redshifts by taking in to account a number of weaker emission lines.
At $3.856 \leq z \leq 4.514$ (z$_{3}$), H$\alpha$ is present in the 3.6$\mu$m band while the 4.5$\mu$m band is contaminated by another potentially strong line [S III] 9068.6 $\mbox{\AA}$. 
Its strength can be estimated from sources in the $2.900 \leq z \leq 3.309$ (z$_{1}$) and $3.416 \leq z \leq 3.829$
(z$_{2}$) redshift intervals in a manner similar to the measurement of H$\alpha$ from z$_{4}$ and z$_{5}$. 
Although [S III] is present in the 3.6$\mu$m band down to z$\approx$2.5, we do not include sources at redshifts lower than z = 2.9. 
The reason is that, unlike most MUSE redshifts at z$\geq$2.9, which are measured from the Lyman-alpha line, MUSE redshifts at z < 2.9 are measured from other spectroscopic features (e.g. O II, C III] and absorption lines), which would make our selection more heterogeneous.  All redshift ranges and the relevant emission lines are also tabulated in Table \ref{table:z_ranges}.

\begin{table*}
	\centering
    \caption[Table caption text]{A list of the different redshift ranges we consider in isolating strong emission lines to specific Spitzer/IRAC bands. }
	\begin{tabular}{cccrr}
		\label{table:z_ranges}
         & \multicolumn{2}{c}{Spitzer/IRAC Band} & \multicolumn{2}{c}{no. of sources} \\
	    redshift range  & [3.6]  & [4.5]  &  MUSE-Deep  &  MUSE-Wide  \\
    	\hline
        \noalign{\smallskip}
	    2.900 < z$_{1}$ < 3.309  &  [S III] 9068.6 \AA  &  &  155  & 131 \\
        \noalign{\smallskip}
        3.416 < z$_{2}$ < 3.829  &  &  [S III] 9068.6 \AA  &  192  & 146 \\
        \noalign{\smallskip}
        3.856 < z$_{3}$ < 4.514  &  H$\alpha$  &  [S III] 9068.6 \AA  &  156  & 164 \\
        \noalign{\smallskip}
        4.532 < z$_{4}$ < 4.955  &  H$\alpha$  &  &  116  & 84 \\
        \noalign{\smallskip}
        5.103 < z$_{5}$ < 5.329  &  &  H$\alpha$  &  36  & 32 \\
        \noalign{\smallskip}
        \hline
	\end{tabular}
\end{table*}

The aforementioned redshift windows are defined only by the major emission lines, i.e. H$\alpha$ and [OIII] 5006.84 $\mbox{\AA}$. 
The [S III] 9068.6 $\mbox{\AA}$ line is also considered to be a major emission line, partially motivated by our observations. 
Weaker lines including [S III] 9530.9 $\mbox{\AA}$, [N II] 6548.05 $\mbox{\AA}$ \& 6583.5 $\mbox{\AA}$, and [S II] 6716.0 $\mbox{\AA}$ \& 6730.0 $\mbox{\AA}$, are considered in our analysis, but are treated as having an amplitude proportional to the stronger lines (see \S4.2). 

\begin{figure*}
  \centering
  {\includegraphics[width=17.2cm]{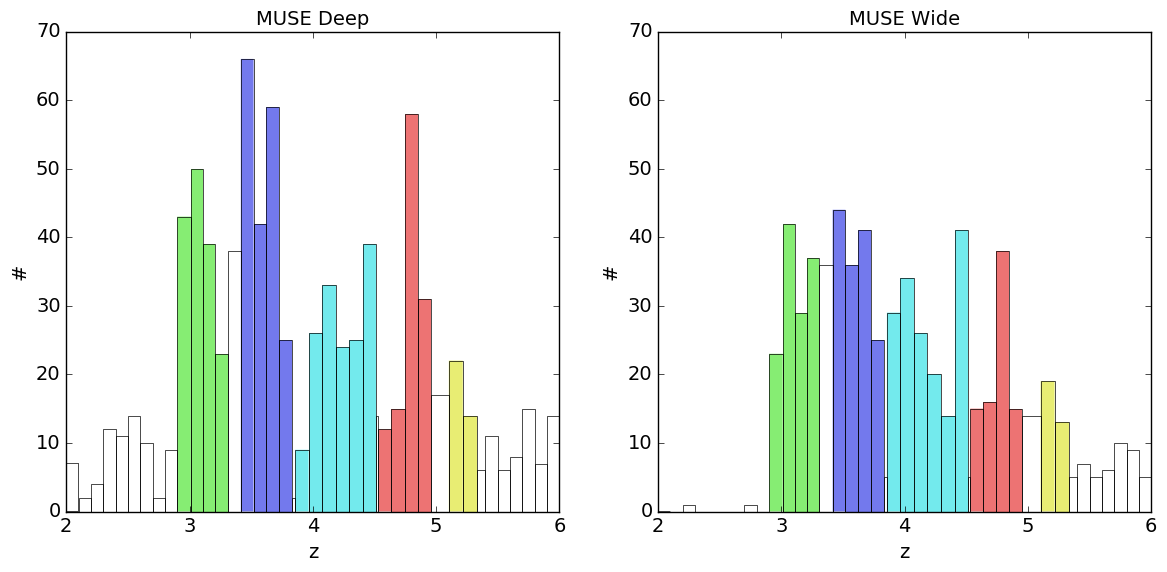}}
  \caption{
Number of galaxies vs. spectroscopic redshift based on our MUSE Deep (\textit{left}) and MUSE wide (\textit{right}) data sets. 
The sources highlighted in green (z$_{1}$) and blue (z$_{2}$) are those where the [S III] 9068.6 $\mbox{\AA}$ line falls in the Spitzer/IRAC 3.6 $\mu$m and in the 4.5 $\mu$m bands, respectively, and no other strong lines seem likely to be present in the other band. 
Sources highlighted in cyan (z$_{3}$) are those where the H$\alpha$ line is present in the 3.6 $\mu$m band and [S III] is in the 4.5 $\mu$m band. 
Sources highlighted in red (z$_{4}$) and yellow (z$_{5}$) are those where H$\alpha$ falls in the 3.6 $\mu$m and the 4.5 $\mu$m bands, respectively, and  no other strong lines are located in the other Spitzer/IRAC band. 
See table \ref{table:z_ranges} for the numbers of sources and figure \ref{fig:halpha} for the line location with respect to the IRAC filters. 
           }
  \label{fig:zdist}
\end{figure*}

Figure \ref{fig:zdist} shows the number of sources with MUSE spectroscopic redshifts that are within our defined redshift windows in MUSE-Deep and MUSE-Wide, respectively. 
From the MUSE-Deep catalog, we obtained 155, 192, 156, 116, and 36 sources in the redshift intervals z$_{1}$, z$_{2}$, z$_{3}$, z$_{4}$, and z$_{5}$, respectively. 
In MUSE-Wide, 131, 146, 164, 84, and 32 sources are identified in these same respective intervals.  Note that the number of usable sources becomes lower when we enforce additional selection criteria, such as whether the spectroscopic redshift is consistent with the photometric redshift, and whether accurate photometry is available.

\subsection{IRAC Observations}

An important aspect to this study is taking advantage of significantly deeper Spitzer/IRAC observations than were available in previous work by Smit et al.\ (2016) and Rasappu et al.\ (2016), as discussed in \S1.

We use data from the GOODS Re-ionization Era wide-Area Treasury from Spitzer (GREATS, PI: Labb\'{e}) combined with all the previous Spitzer/IRAC observations taken over the CDF-South (M. Stefanon et al., in prep.). 
For the 3.6 $\mu$m band, the depth within the area we have MUSE spectroscopy ranges from 76 to 278 hours, with an average depth of 167 hours. 
For the 4.5 $\mu$m band, the depth ranges from 63 to 264 hours, with an average depth of 139 hours. 
The full-width-half-max (FWHM) of the point-spread-function (PSF) of the 3.6 $\mu$m and 4.5 $\mu$m bands are 1.95" and 2.02" respectively. 
By randomly placing 1000 apertures with 0.9" radius on the background in the area where the MUSE sources are located, and measuring the fluxes within the apertures, we find the 5$\sigma$ limiting AB magnitudes in the 3.6 $\mu$m band and the 4.5 $\mu$m band to be 26.6 and 26.5 respectively.

\subsection{HST Observations}

HST observations over our fields allow us to model the SED and provide us with measurements of the $UV$-continuum flux. 
For our HST observations over the fields where we have MUSE spectroscopy, we make use of the Hubble Legacy Field \citep[HLF, ][]{illingworth16} reduction.
Combining observations from 31 observation programs, this HLF reduction constitutes the deepest reduction of the optical and near-infrared data over the GOODS-S region. 
We use the version 1.5 data\footnote{https://archive.stsci.edu/prepds/hlf/} which include the nine filters F435W, F606W, F775W, F814W, F850LP, F105W, F125W, F140W, and F160W. 
F098M is not utilized due to its insignificant depth over the area where we have MUSE redshift coverage.  
Limiting AB magnitudes are measured in the same manner as we did for IRAC data, except using smaller apertures of 0.35" radius. 
For the typical region within the HLF, the 5$\sigma$ limiting AB magnitudes we found for F435W, F606W, F775W, F814W, F850LP, F105W, F125W, F140W, and F160W are 28.08, 27.54, 28.11, 26.73, 28.28, 26.59, 26.51, 25.46, and 26.61, respectively. 
The images we use for our analysis all have a pixel scale of 60 milliarcseconds.  

\section{Photometry and Parameter Inferences}

\subsection{HST Photometry}

Our procedure for performing HST photometry can be briefly described as follows.
We use the original, non-PSF-matched F850LP images as our detection images.
Skipping the PSF-matching step keeps the signal-to-noise at a maximum. 
Unlike most other studies on high-z galaxy observations, we do not construct our detection image from multiple filters. 
A combined detection image has boosted S/N but also has slightly different PSF shapes between stars and galaxies due to their color difference. 
As we will show in section \ref{sec:irac_photometry}, the HST detection image has to be used also as the `prior image' for IRAC photometry. 
Since our sample is limited by detectability in the shallower IRAC data, the benefits of a stable PSF shape are arguably of similar or greater importance to that of an increased S/N in the detection image. 
We then measure the flux ratios of sources using \texttt{SExtractor} in dual-image mode with the aforementioned detection image and each science image that \textit{are} PSF-matched using the procedure we describe in appendix \ref{apdx:psf}.  
We experimented with various combinations of \texttt{SExtractor} parameters and arrived at a set that is optimized for detecting faint and small sources. 
The optimal parameters are listed in Table \ref{table:sex_param}. 
We use the fluxes measured in isophotal apertures as our best estimate of the flux ratios as these apertures give us measurements whose S/N is typically as high or higher than other aperture choices, such as circular and Kron \citep{kron80}. 
We then correct all isophotal fluxes for light that falls beyond the isophotal apertures by multiplying with the factor AUTO$_{\textrm{F850LP}}$/ISO$_{\textrm{F850LP}}$, where AUTO$_{\textrm{F850LP}}$ is the F850LP flux measured by the Kron aperture. 
We correct all uncertainties for noise correlation using equation A20 in \cite{casertano00}, 
\begin{equation}
\sqrt{F_{A}} = 
\begin{cases}
(s/p) (1-\frac{1}{3}s/p) & \textrm{if $s < p$, }\\
1-\frac{1}{3}p/s & \textrm{if $p < s$, }
\end{cases}
\end{equation}
where $F_{A}$ is the factor which divides the aperture-corrected photometry to give the noise correlation-corrected photometry, $s$ the pixel size, and $p$ the drizzle pixfrac parameter. 

We cross-match MUSE sources with HST sources by requiring the presence of a unique HST counterpart within distances of 0.2" and 0.4" for MUSE-Deep and MUSE-Wide respectively. 
We tried cross-matching with maximum allowed offsets ranging from 0.1" to 0.8" in steps of 0.1", and found that the aforementioned values give the highest unique cross-matching rate. 
When the maximum allowed offset is smaller than 0.2"-0.4", we lose sources by scatter in astrometry. 
When the maximum allowed offset is larger than 0.2"-0.4", we find sources in our MUSE catalogues increasingly matching with more than one source in our HST catalogues, implying that at such large matching radii the identifications are more ambiguous.  
Our cross-matching rate in MUSE-Deep of 85\% is comparable with the fraction of MUSE-Deep sources that have HST counterparts in the UVUDF catalog \citep{rafelski15}, which is 89\% \citep{inami17}. 

For MUSE-Wide, our overall cross-matching rate of 76\% is slightly lower those cross-matching with the catalogs of \cite{guo13} and \cite{skelton14}, which are 80\% and 85\% respectively \citep{urrutia18}. 
The deficit is likely caused by the fact that our HST detection scheme is set up for finding high-redshift, faint objects. 
If we only consider MUSE-Wide objects that have spectroscopic redshifts greater than z = 2.9 (the lowest redshift relevant to this work), then we get a much more consistent cross-matching rate of 55\% compared to the 44\% and 57\% cross-matching rate with \cite{guo13} and \cite{skelton14} respectively \citep{urrutia18}. 

\subsection{Spitzer/IRAC Photometry}\label{sec:irac_photometry}

\begin{figure}
\resizebox{\hsize}{!}{\includegraphics{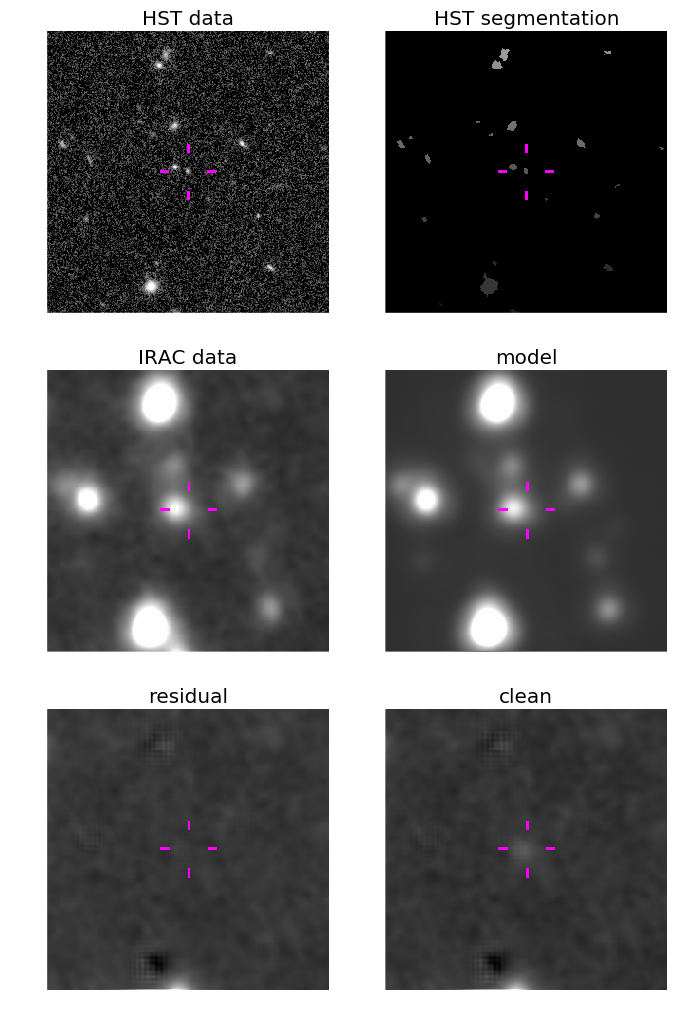}}
\caption{Illustration of how the subtraction of flux from neighboring sources is performed on the Spitzer/IRAC images to derive accurate photometry for one of our sources in our sample.
The `HST data' panel (\textit{upper left}) shows a cut-out of the original F850LP image centered on a galaxy at z=4.54. 
The `HST segmentation' panel (\textit{upper right}) shows the SExtractor segmentation map over the same region. 
The `IRAC data' panel (\textit{middle left}) shows the original \textit{Spitzer/IRAC} 3.6 $\mu$m image. 
The `model' panel (\textit{middle right}) shows the best-fit model constructed from segments of psf-matched \textit{HST} F850LP data. 
The `residual' panel (\textit{lower left}) is the residual subtracting the full model from the data. 
In the `clean' panel (\textit{lower right}), all objects are subtracted except the target itself. 
The \textit{Spitzer/IRAC} 3.6 $\mu$m flux is then measured from the `clean' image. 
Each tile size is 18"$\times$18". }
\label{fig:mophongo}
\end{figure}

Due to the broad PSF of Spitzer/IRAC (FWHM $\approx$2") and the depth of the GREATS data, source crowding makes it difficult for basic photometric tools (e.g. \texttt{SExtractor}) to make accurate flux measurements for faint sources.  Given these challenges, we use the code `MOPHONGO' \citep{labbe15} to model and subtract neighboring sources in the IRAC images within a 12" $\times$ 12" region surrounding the source. 
The source morphology in the IRAC images is modeled using PSF-matched HST images, with the total flux of each source being a free parameter. 
Figure \ref{fig:mophongo} shows an example of the subtraction we obtain of neighboring objects near one of the sources in our samples with a MUSE redshift.

\subsection{Absolute UV magnitudes}
\begin{figure}
\centering
\resizebox{\hsize}{!}{\includegraphics{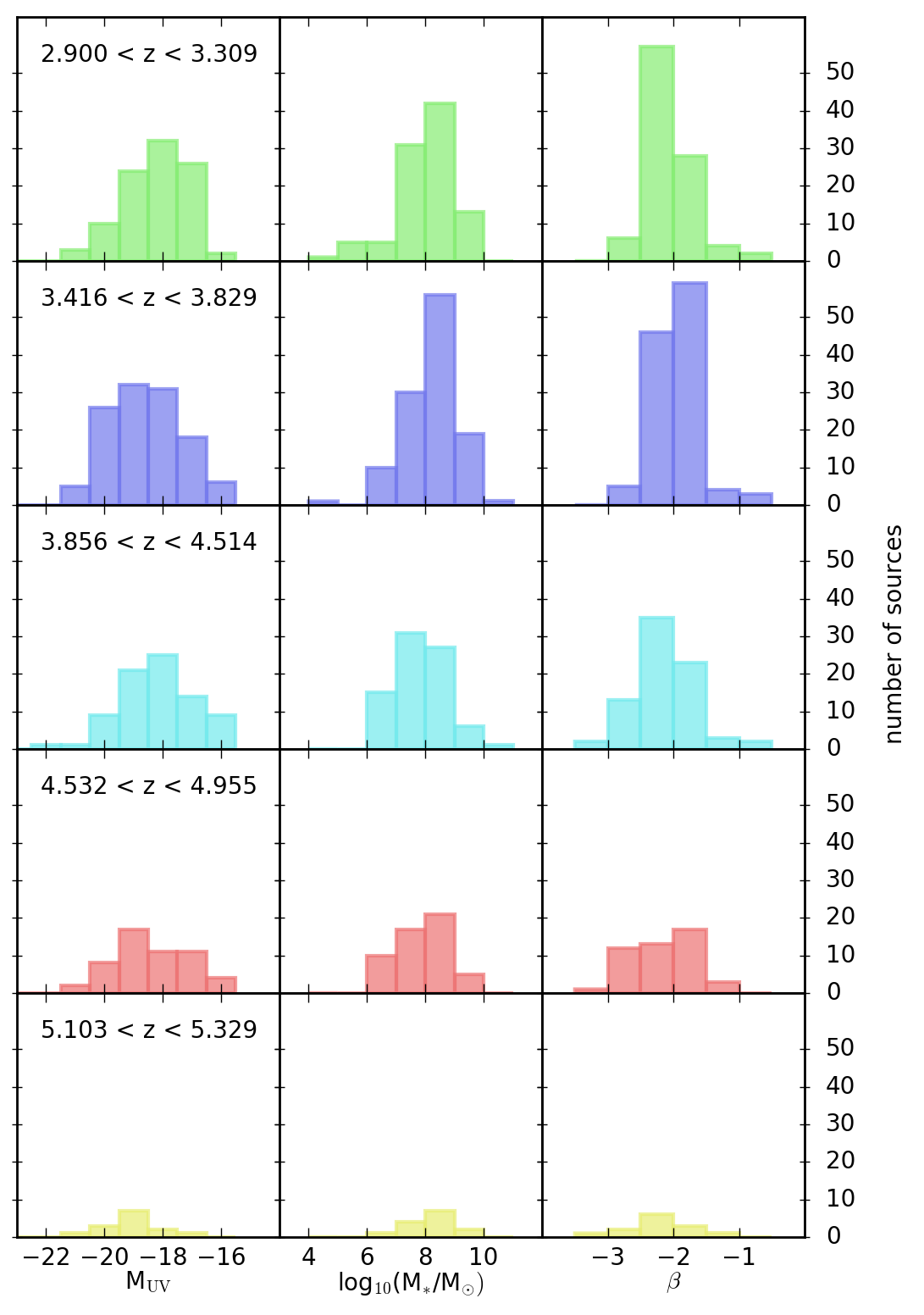}}
\caption{Number of sources in each of the spectroscopic redshift intervals as a function of their absolute magnitudes (\textit{left column}), stellar masses (\textit{middle column}), and UV continuum slopes (\textit{right column}).  In estimating the strength of the H$\alpha$ and [SIII] lines, we make use of stacks of the Spitzer/IRAC images of the sources from each bin.}
\label{fig:histograms}
\end{figure}

The absolute UV magnitude, M$_{UV}$, is a common metric to quantify the intrinsic luminosity of a source. 
It allows us to bin objects with similar intrinsic luminosities to obtain meaningful stacked photometry. 
Following the convention of B16, we measure M$_{UV}$ at a rest-frame wavelength of 1600 $\mbox{\AA}$. 
They are computed from the apparent magnitudes in F606W, F775W, F850LP, and F105W for sources in $2.900 < z < 3.429$, $3.429 < z < 4.250$, $4.250 < z < 5.063$, and $5.063 < z < 5.329$ respectively. 
We correct these magnitudes for the shapes of the transmission curves, and in some cases, the presence of Lyman break (at rest-frame 1216 \r{A}) inside the relevant filter by assuming a simple $f_{\lambda} \propto \lambda^{-2}$ continuum with a sharp Lyman-break cutoff. 
The left column of figure \ref{fig:histograms} shows the distribution of M$_{UV}$ at different redshifts. 

\subsection{Stellar mass inferences}

It is interesting to examine how the emission line strengths and Lyman-continuum photon production efficiency, $\xi_{\textrm{ion}}$, depend on the stellar mass. 
As such, we estimate stellar masses for individual sources to segregate sources into different stellar mass bins. 
The stellar masses are estimated by fitting synthetic stellar population spectra to the HST+Spitzer photometry with FAST \citep{kriek09}. 
We adopt the stellar population synthesis libraries of \citet{conroy10}, a \cite{chabrier03} initial mass function, a delayed exponentially declining star formation history, and a \citet{kriek13} dust law. 
Metallicity is allowed to take values of 0.019 ($Z_{\odot}$), 0.0096 ($\approx Z_{\odot}$/2), and 0.0031 ($\approx Z_{\odot}$/6). 
Redshift is left as a free parameter because the `photometric redshifts' derived by FAST can be used to strengthen the less-confident spectroscopic redshifts (more on this in section \ref{sec:photoz}). 
Figure \ref{fig:histograms} shows the distributions of derived stellar masses. 

At present, FAST is only capable of modeling the continuum radiation without including contributions from the emission lines.   As a result, FAST will try to reproduce the flux in H$\alpha$-boosted bands using features in the continuum.  This could result in  unphysical estimates of the stellar mass. 
It is useful, therefore, in deriving reliable stellar masses to exclude the Spitzer bands which are contaminated by the H$\alpha$ emission line, i.e. the 3.6$\mu$m band for sources in the redshift range 3.856 < z < 4.955, and the 4.5$\mu$m band for sources in the range 5.103 < z < 5.329. 

\subsection{$UV$-continuum slope inferences}

We will also want to examine whether the emission line strengths show a dependence on the $UV$-continuum slope of individual sources -- given that the $UV$-continuum slope $\beta$ is frequently related to the average light-weighted age of the stellar populations within a galaxy.  The $UV$-continuum slope is parameterized as $f_{\lambda} \propto \lambda^{\beta}$. 
Following the convention of \citet{bouwens12} and \citet{castellano12}, we measure $\beta$ by modeling the photometry of HST filters that correspond to rest-frame wavelengths from 1650 to 2300 $\mbox{\AA}$ as a power law.
Table \ref{table:beta_filters} from Appendix C lists the filters used for deriving $\beta$ for sources at different redshifts. 
Figure \ref{fig:histograms} illustrates the distribution of UV continuum slopes derived for our samples. 

\subsection{Photometric redshift inferences}\label{sec:photoz}

As a check on the spectroscopic redshifts inferred from the MUSE GTO data, we also derive photometric redshifts for all the sources. 
The value of doing these checks is clear for the less-certain MUSE redshifts, which are derived from single, low-S/N emission lines, where an asymmetric Ly-$\alpha$-like profile is not evident.
However, the same check is needed even for the more-secure MUSE redshifts (those with high-S/N asymmetric profile of Ly$\alpha$ and/or multiple spectroscopic features). 
The reason is that some of these secure MUSE sources overlap with brighter, foreground objects which could be confused using our cross-matching procedure. 
Clearly, such cases have to be excluded from our analysis because the spectroscopic redshifts and the photometry correspond to different objects. 
We check the spectroscopic redshifts with the photometric redshifts derived with either EAZY \citep{brammer08} or with FAST. 
We define consistency between spectroscopic redshifts $z_{spec}$ and photometric redshifts $z_{phot}$ to be such that $|\Delta (z_{phot} - z_{spec})| < 0.5$.

We derive separate photometric redshifts for each source with FAST and EAZY and only require that one of the two redshift estimates is consistent with a $\Delta z$ $<$ 0.5 for maximum inclusivity.  Fortunately, the two codes give similar results for most sources, despite modest differences in the approaches they utilize.

EAZY derives photometric redshifts by fitting spectral templates, with the flux in one of the bands acting as prior. 
Although the templates used in EAZY contain emission lines, the line strengths are not free parameters.  Since this may have a significant impact on the ability to fit fluxes in bands with H$\alpha$ line contamination, we excluded those bands from the fits. 

After excluding sources with unsatisfactory neighbor-subtraction in the IRAC images (see section \ref{sec:stack} for the exact criteria), we found that 77\% of the sources with less-confident MUSE redshifts are consistent with our photometric redshift estimates.
For the 219 sources with confident MUSE redshifts, only two are `incompatible' with our photometric redshift estimates. 
In both cases, this appears to be due to overlap with other foreground sources. 

\section{Empirical Estimate of $\xi_{\textrm{ion}}$} 

\subsection{Stacking the IRAC and HST Images}\label{sec:stack}

For galaxies in each of our subsamples (by $UV$ luminosity, stellar mass, and $\beta$), we derive a weighted average flux in each IRAC band. 
The weight scheme is designed to both maximize the S/N and determine the representative SED.  
To this end, we stack the IRAC images, after all the neighboring objects have been subtracted, i.e., the `clean' panel in Figure~\ref{fig:mophongo}.  To optimize the accuracy of the flux measurements we obtain from the stacks, only images with a satisfactory subtraction of their neighbors are considered for inclusion. 
Quantitatively, we exclude any sources which, in either the 3.6 $\mu$m band or the 4.5 $\mu$m band, have a reduced chi-squared larger than 1.4, flux uncertainty larger than 30.0 nJy, or contamination by neighbors higher than 550\% of the measured F850LP flux. 
These criteria are set in order to maximize the usable sample size, and are determined by visual inspection of the subset of galaxies at $5.103 < z < 5.329$. 

In addition to these cutoffs, each source contributes to the final mean stack based on a three-component weight.  Each weight is composed of a reduced chi-squared component, an uncertainty component, and a contamination fraction component,
\begin{equation}
\label{eq:weight_components}
w = w_{\chi^{2}} + w_{\textrm{err}} + w_{\textrm{cont}} ,
\end{equation}
where each component is parametrized as follows:
\begin{equation}
\label{eq:weight_scheme}
\begin{split}
w_{\chi^{2}} &=
\begin{cases}
1 & \mathrm{(0.0 < \chi^{2} < 0.02)} \\
\frac{1.4-\chi^{2}}{1.4-0.02} & \mathrm{(0.02 < \chi^{2} < 1.4)} \\
0 & \mathrm{(\chi^{2} > 1.4)}
\end{cases} \\
w_{\textrm{err}} &=
\begin{cases}
1 & \mathrm{(0.0\,nJy < err < 10.0\,nJy)} \\
\frac{30.0-err}{30.0-10.0} & \mathrm{(10.0\,nJy < err < 30.0\,nJy)} \\
0 & \mathrm{(err > 30.0\,nJy)}
\end{cases} \\
w_{\textrm{cont}} &=
\begin{cases}
1 & \mathrm{(0.0 < cont < 1.0)} \\
\frac{5.5-cont}{5.5-1.0} & \mathrm{(1.0 < cont < 5.5)} \\
0 & \mathrm{(cont > 5.5)}
\end{cases}
\end{split}
\end{equation}
We examine three different binning schemes in examining the dependencies of $\xi_{\textrm{ion}}$ on various physical or observational quantities. 
First, sources are binned according to their absolute UV magnitudes: $-21.5 < \textrm{M}_{UV} < -20.5$, $-20.5 < \textrm{M}_{UV} < -19.5$, $-19.5 < \textrm{M}_{UV} < -18.5$, $-18.5 < \textrm{M}_{UV} < -17.5$. 
Second, the binning is done according to the stellar mass: $10.0 > \log_{10} \textrm{M}_{*}/\textrm{M}_{\odot} > 9.0$, $9.0 > \log_{10} \textrm{M}_{*}/\textrm{M}_{\odot} > 8.0$, and $8.0 > \log_{10} \textrm{M}_{*}/\textrm{M}_{\odot} > 7.0$. 
Third, the binning is performed according to the UV continuum slope: $-1.5 > \beta > -2.0$, $-2.0 > \beta > -2.5$, and $-2.5 > \beta > -3.1$. 
The distribution of weights $w$ used for the $M_{UV}$, $M_{*}$, and $\beta$-subsample stacks is presented in figure \ref{fig:weights} from Appendix D. 

Of course, for the deep stacks we create for galaxies in different bins of stellar mass M$_{*}$ and $\beta$, we include sources over a very wide range in $UV$ luminosity.  For most of these stacks, the brightest and faintest sources typically differ in luminosity by a factor of $\approx$10, with the most extreme case occurring in a bin with $-$2.0$>\beta>$$-$2.5 and $3.105<z<3.309$.  There the brightest source is 1258$\times$ more luminous than the faintest source.  In that case, the brightest source receives 35\% of the total weight for that bin.  For the other 18 sources which make up the bin, the contribution is $\approx$4\%.

\begin{figure*}[t]
  \centering
  {\includegraphics[width=130mm]{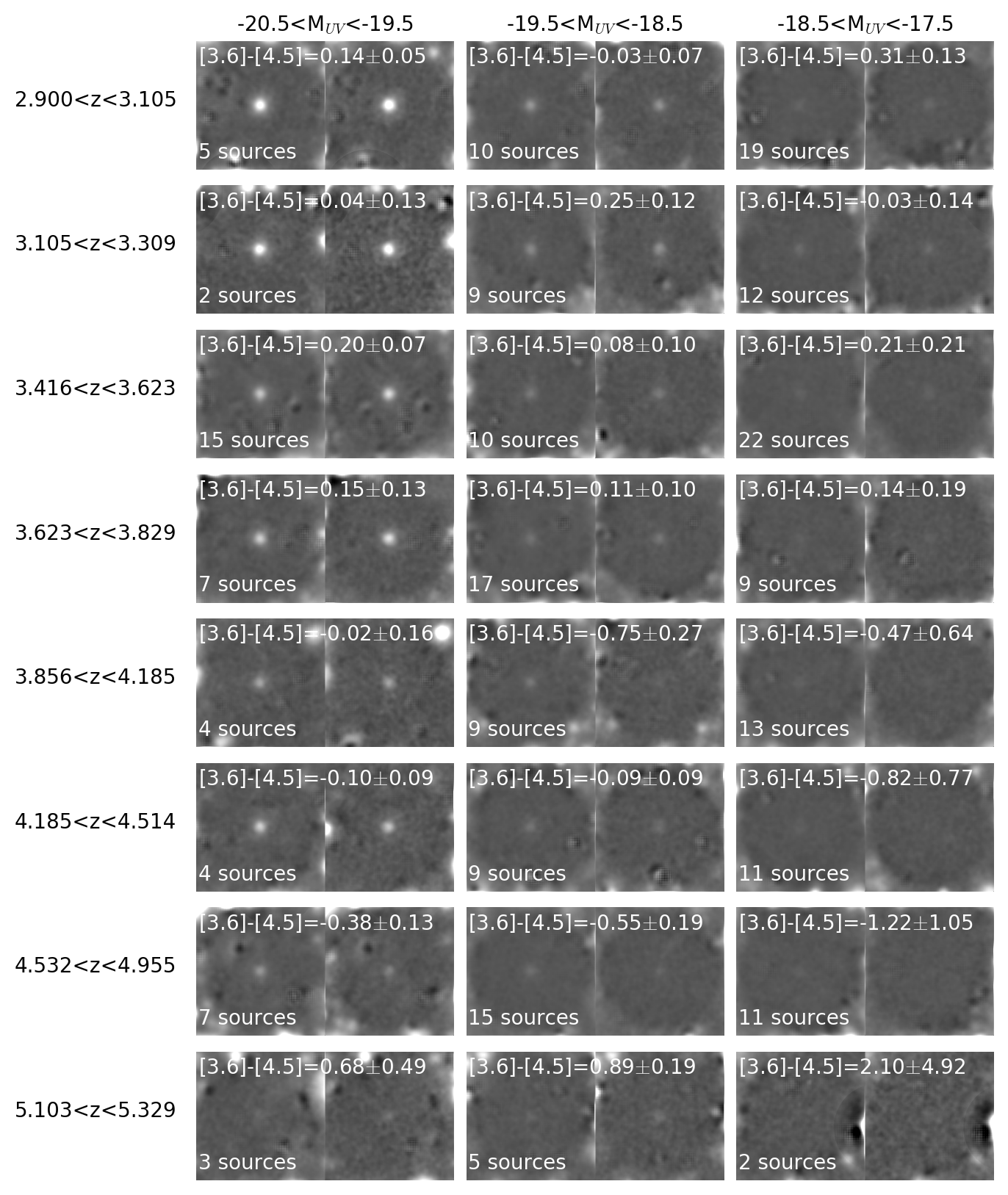}}
  \caption{Stacked Spitzer/IRAC [3.6] (left) and [4.5] (right) images of sources found in the redshift intervals 2.900 < $z$ < 3.105, 3.105 < $z$ < 3.309, 3.416 < $z$ < 3.623, 3.623 < $z$ < 3.829, 3.856 < $z$ < 4.185, 4.185 < $z$ < 4.514, 4.532 < $z$ < 4.955, and 5.105 < $z$ < 5.329.
    Only sources with satisfactory neighbor subtractions and spectroscopic redshifts consistent with their photometric redshifts are included in the stacks.
    The stacks are binned into three groups of intrinsic luminosity,  $-20.5 < \textrm{M}_{UV} < -19.5$, $-19.5 < \textrm{M}_{UV} < -18.5$, and $-18.5 < \textrm{M}_{UV} < -17.5$. 
    Some neighboring objects remain in the corners because the subtraction is performed within a diameter of 12" instead of a square of the same size. }
  \label{fig:stack_MUV}
\end{figure*}

\begin{figure*}[t]
  \centering
  {\includegraphics[width=130mm]{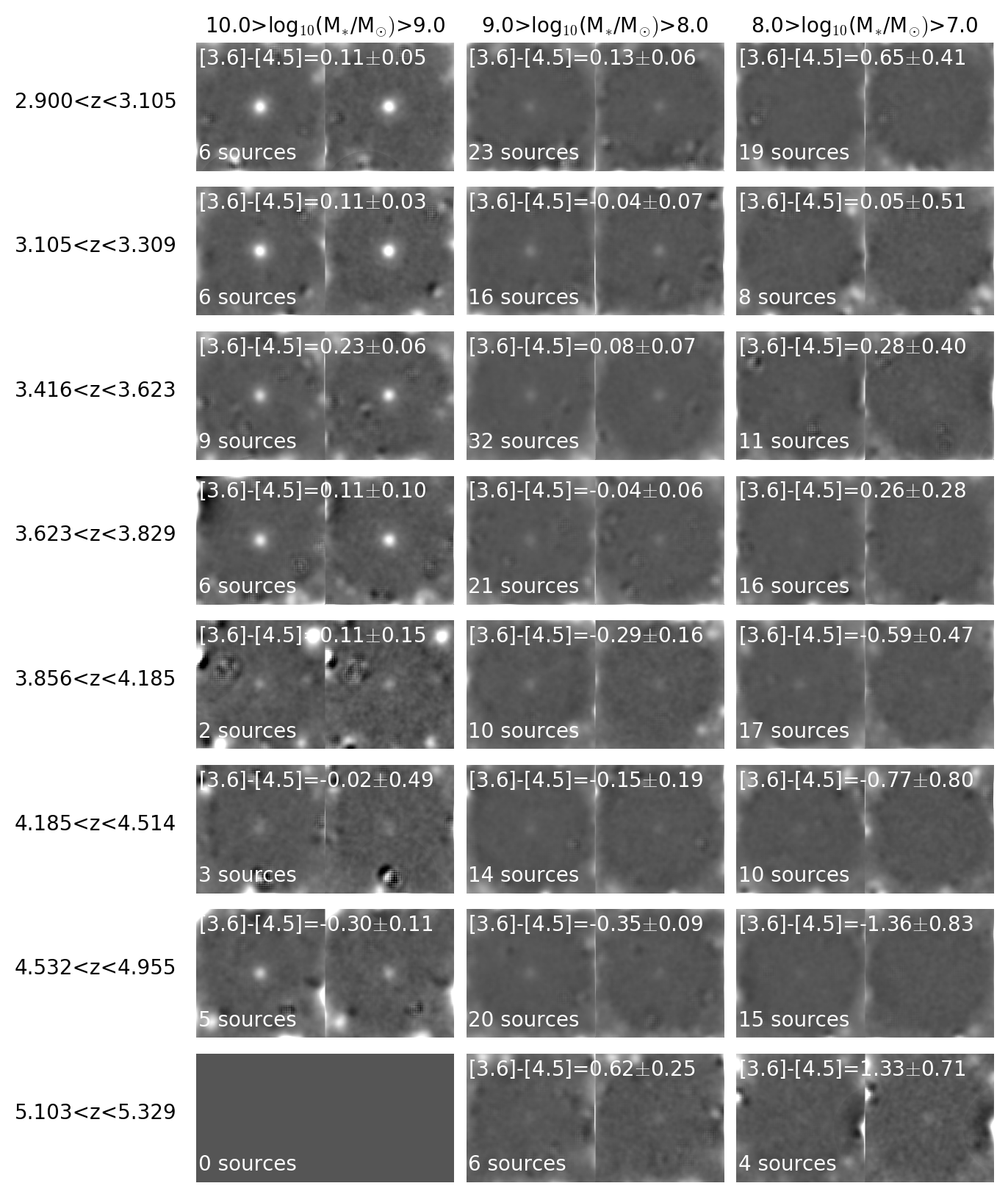}}
  \caption{Stacked Spitzer/IRAC [3.6] (left) and [4.5] (right) images of sources vs. their estimated stellar mass in the same redshift intervals considered in Figure~\ref{fig:stack_MUV}. 
  Three different bins in stellar mass are considered: $10.0 < \log_{10} (\textrm{M}_{*}/\textrm{M}_{\odot}) < 9.0$, $9.0 < \log_{10} (\textrm{M}_{*}/\textrm{M}_{\odot}) < 8.0$, and $8.0 < \log_{10} (\textrm{M}_{*}/\textrm{M}_{\odot}) < 7.0$. }
  \label{fig:stack_mass}
\end{figure*}

\begin{figure*}[t] \centering {\includegraphics[width=130mm]{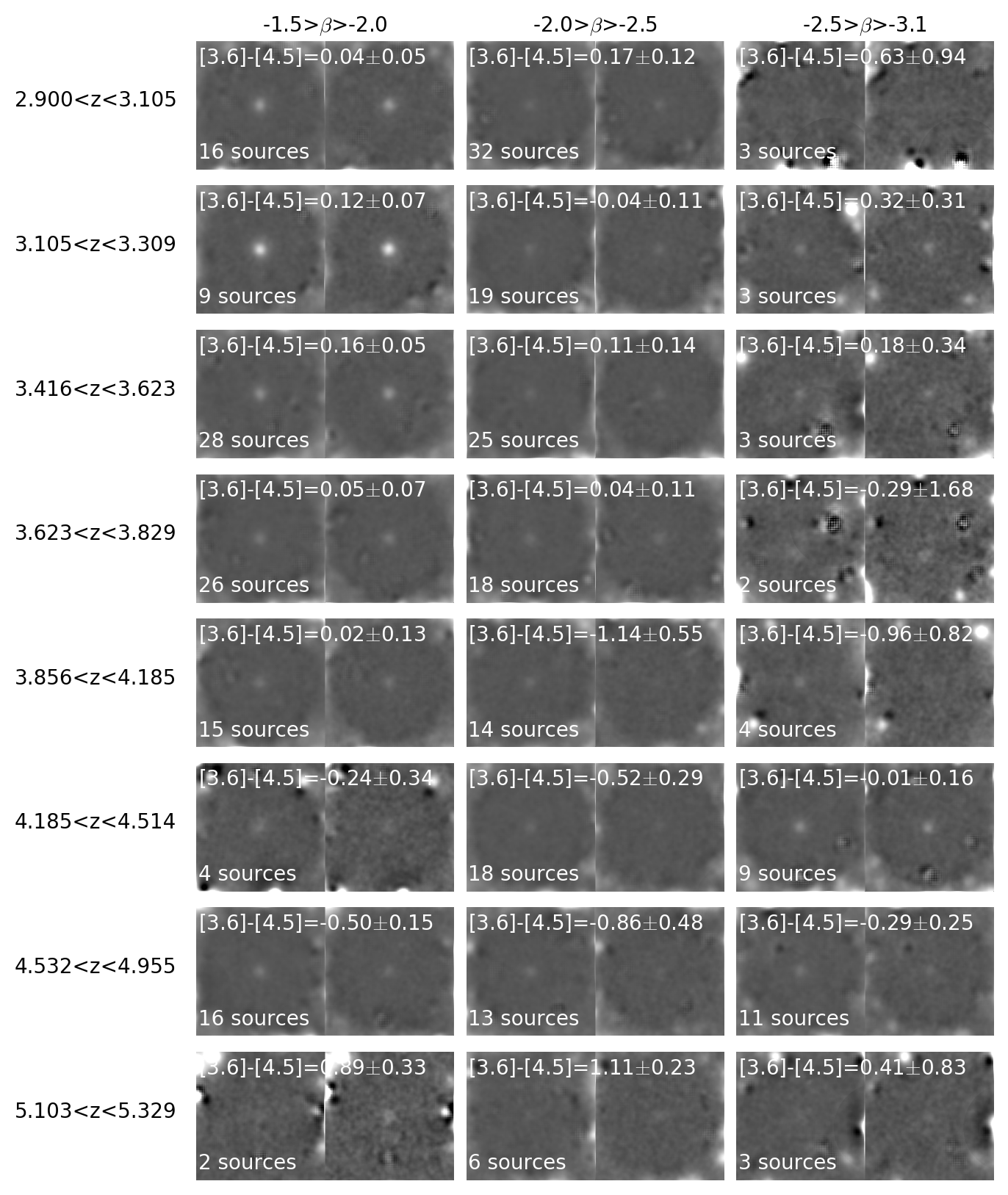}}
 \caption{Stacked Spitzer/IRAC [3.6] (left) and [4.5] (right) images of sources vs. their UV-continuum slope $\beta$ in the same redshift intervals considered in Figure~\ref{fig:stack_MUV}. 
 Three bins in $\beta$ are considered: $-1.5 > \beta > -2.0$, $-2.0 > \beta > -2.5$, and $-2.5 > \beta > -3.1$. }
  \label{fig:stack_beta}
\end{figure*}

In addition, in order to provide a larger number of constraints for our modeling, the first three redshift ranges, z$_{1}$, z$_{2}$, and z$_{3}$, are each divided into two pieces for the purpose of the binned stacks. 
That is, the redshift interval z$_{1}$ is divided into two separate intervals $2.900 < z < 3.105$ and $3.105 < z < 3.309$, redshift interval z$_{2}$ is divided into $3.416 < z < 3.623$ and $3.623 < z < 3.829$, and redshift interval z$_{3}$ is divided into $3.856 < z < 4.185$ and $4.185 < z < 4.514$. 
Figures \ref{fig:stack_MUV}, \ref{fig:stack_mass}, and \ref{fig:stack_beta} show the stacks for M$_{UV}$, stellar mass, and $\beta$ respectively. 

For symmetry, we have applied an identical weighting scheme when considering quantities derived from the HST photometry, i.e., $L_{UV}$ and $\beta$, for individual sources as we used in weighting these sources in our IRAC stacks.  Since our weighting scheme favors sources where the subtraction of neighboring sources is accurate, IRAC-bright objects would not be expected to dominate the stacks.

\subsection{Measurement of H$\alpha$ Equivalent Widths}\label{sec:colors_fit}

To derive the equivalent widths of H$\alpha$, we first measure the fluxes on the stacked `clean' IRAC images using circular apertures with a radius of 0.9". 
The uncertainties are derived by bootstrapping, which accounts for both source-to-source variations and noise correlation. 
Specifically, we randomly draw, with replacement, $n$ times from the bin sample, where $n$ is the size of the bin sample. 
These $n$ sources are then stacked, and the flux measured. 
We repeat this process 1000 times, and the photometric uncertainty is calculated by taking the standard deviation of the 1000 flux values. 

\begin{figure}
  \centering
  \resizebox{\hsize}{!}
   {\includegraphics{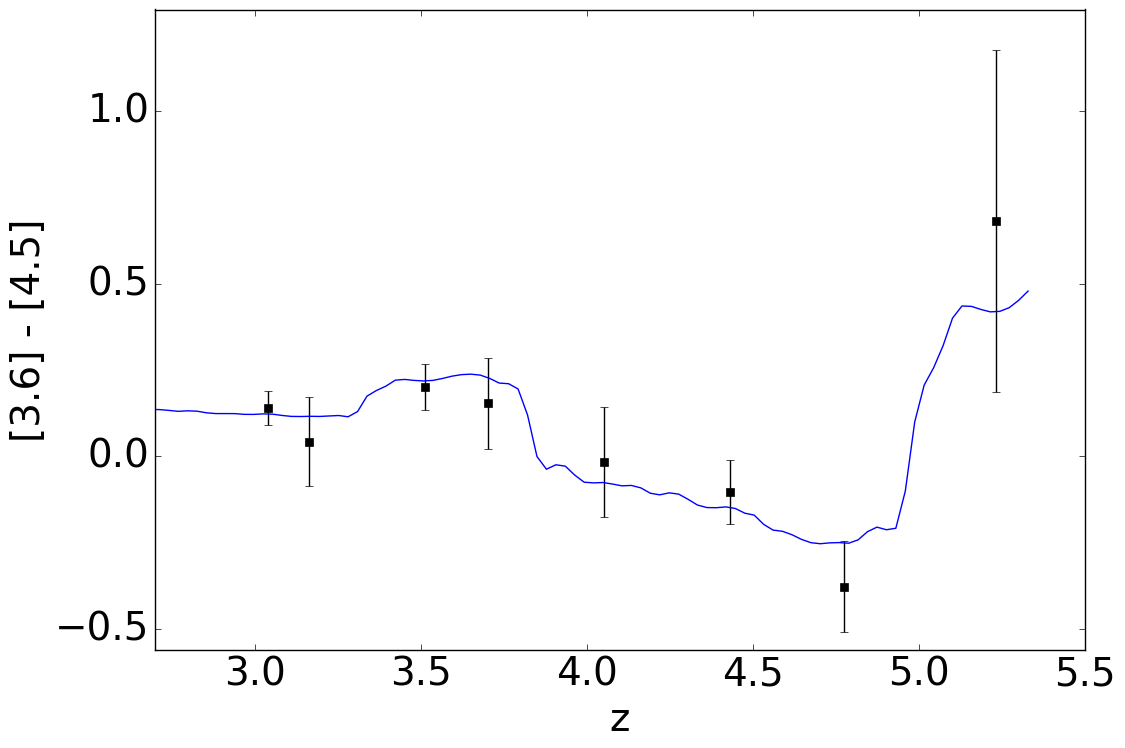}}
  \caption{[3.6]$-$[4.5] color as a function of the redshift bin considered for our stacked Spitzer/IRAC images of sources with $-20.5 < \textrm{M}_{UV} < -19.5$. 
  The best-fit color model is shown in blue. 
  The color model is the convolution of the 3.6 $\mu$m and 4.5 $\mu$m transmission curves 
  with a simple model spectrum, which consists of a power-law continuum, an H$\alpha$ line, 
  an [S III] 9068.6 \r{A} line, and five secondary lines whose strengths are fixed relative to that of the H$\alpha$ line (\S\ref{sec:colors_fit}).} \label{fig:colors_fit}
\end{figure}

We then fit a simple model spectrum, convolved with the filter transmission curves, to the measured [3.6]$-$[4.5] colors as a function of redshift. 
As an example, Figure \ref{fig:colors_fit} shows the measured [3.6 $\mu$m]$-$[4.5 $\mu$m] colors for sources in the $-20.5 < M_{UV} < -19.5$ bin. 
The model spectrum consists of a power-law continuum ($f_{\lambda} \propto \lambda^{\beta_{opt}}$), an H$\alpha$ emission line, an [S III] 9068.6 $\mbox{\AA}$ emission line, and five other secondary emission lines ([N II] 6548.05 $\mbox{\AA}$ \& 6583.5 $\mbox{\AA}$, [S II] 6716.0 $\mbox{\AA}$ \& 6730.0 $\mbox{\AA}$, and [S III] 9530.9 $\mbox{\AA}$) whose strengths are fixed relative to that of H$\alpha$. 
Assuming a metallicity of $Z$ = 0.004 = $Z_{\odot}$/5, we take the strengths of the secondary lines, relative to H$\beta$, and the H$\alpha$/H$\beta$ ratio from \citet{anders03} and \citet{leitherer95}, respectively. 
The non-hydrogen line ratios of \citet{anders03} are based on the modeling results of \citet{stasinska84}, who simulate nebular emission given a wide range of physical conditions. 
\citet{leitherer95} predict the H$\alpha$/H$\beta$ ratio by assuming a gas temperature of 10,000 K, a 10\% helium abundance relative to hydrogen, and case B recombination. 
For simplicity and to avoid overfitting our data, we took the optical continuum slope, the H$\alpha$ EW, and [S III] EW to be independent of redshift. 
The fluxes of the secondary lines range from 2\% to 7\% of the H$\alpha$ flux. 
As an example of how our stacked spectra would look, Figure \ref{fig:muv_stack_spectra} shows the implied spectra for $4.532<z<4.955$ galaxies in three different bins of $UV$ luminosity $M_{UV}$, i.e., $-20.5<M_{UV}<-19.5$, $-19.5<M_{UV}<-18.5$, and $-18.5<M_{UV}<-17.5$.

We find the rest-frame equivalent widths to be $EW_{H\alpha} = 86\pm 15\mbox{\AA}$, $EW_{H\alpha} = 119\pm 52\mbox{\AA}$, and $EW_{H\alpha} = 327\pm 183\mbox{\AA}$ for sources in the bins $-20.5 < \textrm{M}_{UV} < -19.5$, $-19.5 < \textrm{M}_{UV} < -18.5$, and $-18.5 < \textrm{M}_{UV} < -17.5$, respectively. 
Optimal values of all free parameters are listed in Table \ref{table:best_params}. 
Similarly, we infer the H$\alpha$ equivalent widths for M$_{*}$-binned and $\beta$-binned sources, which are also listed in Table \ref{table:best_params}. 

\begin{table*}
  \caption[]{Best-fit values of free parameters and the derived $\xi_{\textrm{ion,0}}$. Rest-frame equivalent widths are calculated assuming $z$ = 3.707 and $z$ = 4.931 for [S III] and H$\alpha$, respectively. }
     \label{table:best_params}
     $$
     \begin{array}{cccccc}
        \hline
         \noalign{\smallskip}
         \textrm{bins}  &  \textrm{no. of sources}  &  \beta_{\textrm{opt}}  &  \textrm{EW}_{\textrm{0, H}\alpha} [\mbox{\AA}]  &  \textrm{EW}_{\textrm{0, [S III]}} [\mbox{\AA}]  &  \textrm{log}_{10} \hspace{0.1cm} \xi_{\textrm{ion,0}} [\mathrm{Hz \hspace{0.1cm} erg}^{-1}]  \\
         \noalign{\smallskip}
         \hline
         \noalign{\smallskip}
         -20.5 < \textrm{M}_{UV} < -19.5  &  47  &  -1.58 \pm 0.12  &  453 \pm 84  &  105 \pm 55  &  25.28^{+0.08}_{-0.09} \\
         \noalign{\smallskip}
         -19.5 < \textrm{M}_{UV} < -18.5  &  84  &  -1.60 \pm 0.35  &  621 \pm 296  &  < 188  &  25.31^{+0.12}_{-0.17}  \\
         \noalign{\smallskip}
         -18.5 < \textrm{M}_{UV} < -17.5  &  99  &  -1.40 \pm 0.32  &  1846 \pm 953  &  < 147  &  25.49^{+0.15}_{-0.22}  \\
         \noalign{\smallskip}
         \hline
         \noalign{\smallskip}
         10.0 < \log_{10} (\textrm{M}_{*}/\textrm{M}_{\odot}) < 9.0  &  37  &  -1.62 \pm 0.11  &  403 \pm 113 &  112 \pm 50 &  25.44^{+0.10}_{-0.12}  \\
         \noalign{\smallskip}
         9.0 < \log_{10} (\textrm{M}_{*}/\textrm{M}_{\odot}) < 8.0   &  142  &  -2.11 \pm 0.16  &  488 \pm 137  &  < 76  &  25.35^{+0.12}_{-0.17}  \\
         \noalign{\smallskip}
         8.0 < \log_{10} (\textrm{M}_{*}/\textrm{M}_{\odot}) < 7.0   &  100  &  -0.56 \pm 0.31  &  2818 \pm 773  &  < 161  &  25.54^{+0.14}_{-0.20}  \\
         \noalign{\smallskip}
         \hline
         \noalign{\smallskip}
         -1.5 > \beta > -2.0  &  116  &  -1.84 \pm 0.16  &  553 \pm 217  &  110 \pm 83  &  25.29^{+0.12}_{-0.16}  \\
         \noalign{\smallskip}
         -2.0 > \beta > -2.5  &  145  &  -1.64 \pm 0.29  &  1767 \pm 501  &  < 156  &  25.62^{+0.14}_{-0.20} \\
         \noalign{\smallskip}
         -2.5 > \beta > -3.1  &  38  &  -1.04 \pm 0.53  &  537 \pm 248  &  < 195  & 25.18^{+0.15}_{-0.22}  \\
         \noalign{\smallskip}
         \hline
      \end{array}
  $$ 
\end{table*}

\subsection{Procedure to Derive $\xi_{\textrm{ion,0}}$}

$\xi_{\textrm{ion}}$ in Equation \ref{eq:definition} refers to the Lyman-continuum photon production efficiency in the presence of a non-zero $f_{\textrm{esc,LyC}}$. 
Since the precise value of $f_{\textrm{esc,LyC}}$ is still uncertain, it has become customary (following B16) to leave out this complication by assuming $f_{\textrm{esc,LyC}}$ = 0, 
\begin{equation}\label{eq:xi_ion0_def}
\xi_{\textrm{ion,0}} \equiv \xi_{\textrm{ion}} \hspace{0.1cm} (1-f_{\textrm{esc,LyC}}) \hspace{0.1cm} \textrm{.}
\end{equation}

The intrinsic production rate of Lyman-continuum photons, which we define as $\dot{N}(H^{0})$, is related to the rate of Lyman-continuum photons reaching the intergalactic medium, $\dot{N}_{\textrm{ion}}$ (capital alphabets denote a change from rate densities to rates), by
\begin{equation}\label{eq:rates}
    \dot{N}_{\textrm{ion}} = \frac{\dot{N}(H^{0}) f_{\textrm{esc,LyC}}}{1-f_{\textrm{esc,LyC}}}
    \textrm{ ,}
\end{equation}
where the $1/(1-f_{\textrm{esc,LyC}})$ factor reflects the fact that $\dot{N}(H^{0})$ is derived from the observed, unattenuated H$\alpha$ flux, which is in turn produced in recombination cascades after unescaped Lyman-continuum photons are absorbed by the neutral hydrogen gas in galaxies. 

Using quantum mechanical simulations, \citet{leitherer95} found the following relation between the H$\alpha$ luminosity and the intrinsic Lyman-continuum photons production rate:
\begin{equation}\label{eq:ha_ion_photon}
L(H\alpha) [\textrm{erg s}^{-1}] = 1.36 \times 10^{-12} \dot{N}(H^{0}) [s^{-1}] \textrm{.}
\end{equation}

We calculate the H$\alpha$ flux, $L(H\alpha)$, using the equivalent widths and rest-frame optical slopes derived in section \ref{sec:colors_fit}. 
The continuum in the H$\alpha$-boosted band is calculated by extrapolating from the photometry in the line-free band using the best-fit rest-frame optical slope. 
The H$\alpha$ flux is corrected for the expected dust extinction assuming an SMC extinction law, which is $(f_{\textrm{esc,UV}})^{2.6/13.2}$. 
Our choice for this particular dust law is motivated by the recent ALMA finding that $z \approx 5-6$ galaxies show resemblance to the SMC dust law \citep{capak15,bouwens16c}, particularly at lower luminosities \citep{bouwens16c}. 

Substituting Equation \ref{eq:xi_ion0_def} and \ref{eq:rates} into \ref{eq:definition}, we can express $\xi_{\textrm{ion,0}}$ as the ratio of the Lyman-continuum photons production rate to the unattenuated rest-frame UV luminosity, 
\begin{equation}\label{eq:xi_ion}
\xi_{ion,0} = \frac{\dot{N}(H^{0})}{L_{UV}/f_{esc,UV}} \hspace{0.1cm} \mathrm{, }
\end{equation}
The observed rest-frame $UV$ luminosity, L$_{UV}$, is calculated from $M_{UV}$ corrected by an escape fraction for UV photons, $f_{\textrm{esc,UV}}$, defined by an SMC dust law \citep{prevot84}, 
\begin{equation}
\label{eq:smc}
f_{\textrm{esc,UV}} =
\begin{cases}
10^{1.1(\beta+2.23)/-2.5} & \textrm{for $\beta > -2.23$} \\
1 & \textrm{for $\beta \leq -2.23$ .}
\end{cases}
\end{equation}
If instead we adopt the \cite{calzetti00} dust law, we recover $\xi_{\textrm{ion,0}}$ values that are lower by $\approx$0.04 dex.

\subsection{$\xi_{ion,0}$ vs. $\textrm{M}_{UV}$, $\textrm{M}_{*}$, and $\beta$\label{sec:xiion}}

For each $\textrm{M}_{UV}$, $M_{*}$, and $\beta$ bin, we obtain an optimized value of the H$\alpha$ EW by fitting the [3.6]$-$[4.5] colors with respect to redshift (e.g., Figure \ref{fig:colors_fit}). 
While by construction, the $L_{UV}$'s of galaxies are fairly uniform across redshift in $\textrm{M}_{UV}$ bins, larger variations exist in the $\textrm{M}_{*}$ and $\beta$ bins. 
Therefore, we calculate $\xi_{\textrm{ion,0}}$ for each relevant redshift (only z$_{3}$, z$_{4}$, and z$_{5}$, where H$\alpha$ is imaged in either IRAC bands) and physical property bin, using the same H$\alpha$ EW derived for that physical property bin, and a distinct $L_{UV}$ for that redshift bin. 
The values and uncertainties of $\xi_{\textrm{ion,0}}$ averaged over redshift are listed in Table \ref{table:best_params}. 
These average values are calculated by weighting each $\xi_{\textrm{ion,0}}$ value by the inverse of their uncertainty squared. 

\begin{figure*}
  \centering
  \resizebox{\hsize}{!}
  {\includegraphics{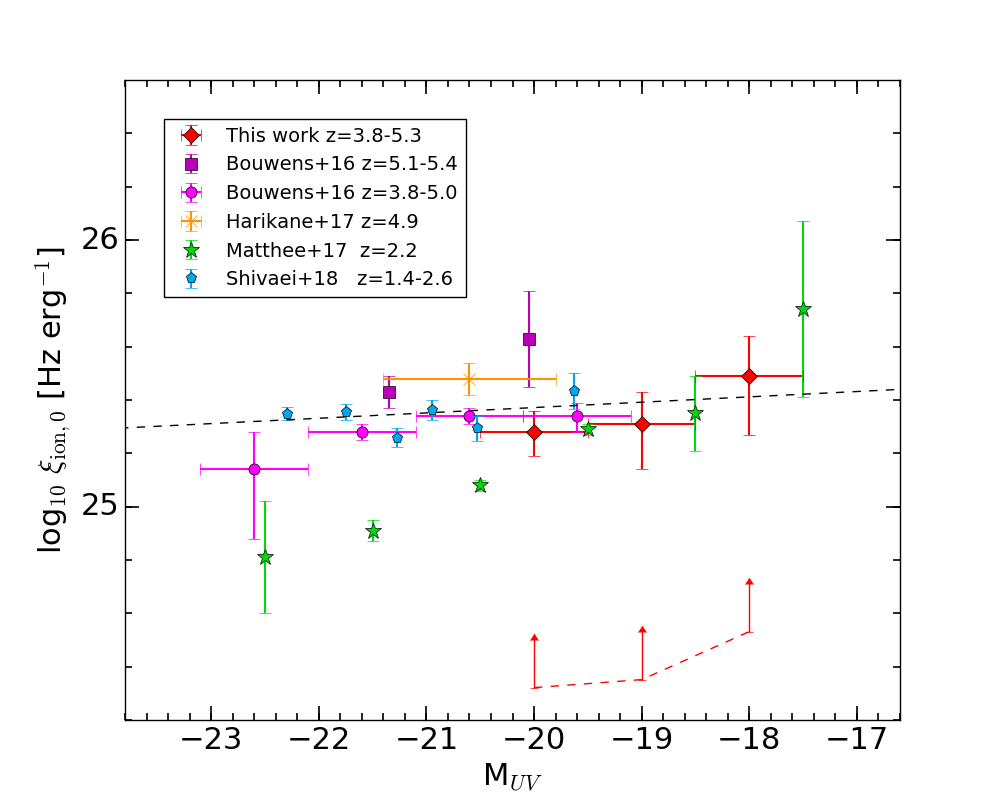}}
  \caption{The present estimates of Lyman continuum photon production efficiency $\xi_{\textrm{ion,0}}$ vs. absolute $UV$ luminosity M$_{UV}$.  
  The red dashed line denotes the lower limit for the population-averaged $\xi_{\textrm{ion,0}}$ obtained by making the extreme assumption that all sources in UV selections not appearing in our MUSE selections have $\xi_{\textrm{ion,0}}$ = 0 (see \S\ref{sec:xiion} for details).  
  For comparison, also shown here are several previous estimates of $\xi_{\textrm{ion,0}}$ by \citet{bouwens16} at $z=3.8$-5.0 (\textit{light magenta}) and $z=5.1$-5.4 (\textit{dark magenta}; using a revised binning, see section \ref{sec:prevwork}.), \citet{harikane18} at z=4.9 (\textit{orange}), \citet{matthee17} at z=2.2 (\citet{meurer99} $\beta$-dust correction, \textit{green}), and \citet{shivaei18} at z=1.4-2.6 (SMC dust correction, \textit{blue}). The black dashed line denotes the best-fit $M_{UV}$ dependence of $\xi_{\textrm{ion,0}}$ on $M_{UV}$ using the measurements of \citet{bouwens16} and this work.  The best-fit relation is $\xi_{\textrm{ion,0}} = 0.020 (\pm 0.031) ($M$_{UV}+20) + 25.372 (\pm 0.045)$.}
  \label{fig:xi_vs_muv}
\end{figure*}

Figure \ref{fig:xi_vs_muv} plots the derived $\xi_{\textrm{ion,0}}$ against M$_{UV}$. 
The inferred values of log$_{10}(\xi_{\textrm{ion,0}} / Hz erg^{-1})$ are 25.28$^{+0.08}_{-0.09}$, 25.31$^{+0.12}_{-0.17}$, and 25.49$^{+0.15}_{-0.22}$ Hz erg$^{-1}$ for $-$20.5<$M_{UV}$<$-$19.5, $-$19.5<$M_{UV}$<$-$18.5, and $-$18.5<$M_{UV}$<$-$17.5, respectively. 
Given that our selections exclusively include sources with a detectable Ly$\alpha$ emission line, it is possible that our selections to show a higher $\xi_{\textrm{ion}}$ than the typical star-forming galaxy at $z\sim3$-6 (see \S\ref{sec:prevwork}). 
We can set a firm lower limit on the population-averaged $\xi_{\textrm{ion,0}}$ by calculating its value in each M$_{UV}$ bin and then assuming galaxies that are detected in HST images but not by MUSE do not contribute at all to the ionizing photon budget (clearly an extreme assumption).  By cross-matching the combined MUSE-Deep and MUSE-Wide catalog with our HST catalog, we found that 11\%, 16\%, and 16\% of the HST-detected sources are detected by MUSE in $-$20.5<$M_{UV}$<$-$19.5, $-$19.5<$M_{UV}$<$-$18.5, and $-$18.5<$M_{UV}$<$-$17.5, respectively. 
These Ly$\alpha$ fractions imply firm lower limits on population-averaged $\log_{10}\,\xi_{\textrm{ion,0}}$'s of 24.32, 24.35, and 24.53 for the aforementioned M$_{UV}$ ranges respectively.

While the redshift evolution of $\xi_{\textrm{ion,0}}$ is an important topic, we do not consider it in this paper due to our moderate number of sources.
The lack of a large sample over a range of redshifts forces us to adopt a constant H$\alpha$ EW model, and so the derived evolution of $\xi_{\textrm{ion,0}}$ is not very meaningful and only reflects the changes in $L_{UV}$. 

\section{Discussion}

\subsection{Comparison with Previous Studies\label{sec:prevwork}}

In this paper, we derived $\xi_{\textrm{ion,0}}$ from the H$\alpha$ equivalent widths measured in broadband Spitzer/IRAC images. 
This approach was also taken by two previous studies \citep{bouwens16, harikane18}. 
Comparing with previous work, we find similar values of $\xi_{\textrm{ion,0}}$ in $-20.5 < M_{UV} < -18.5$ galaxies, and that in the faintest bin ($-18.5 < M_{UV} < -17.5$) is perhaps $\approx$0.2 dex higher, but that increase is not significant.  Figure \ref{fig:xi_vs_muv} plots these $\xi_{\textrm{ion,0}}$ values with respect to M$_{UV}$. 
The spectroscopic sample of \citet{bouwens16} is from \citet{smit16} and \citet{rasappu16}, which, in turn, derive from \citet{ando04}, \citet{stark13}, \citet{stark17}, \citet{balestra10}, \citet{vanzella05}, \citet{vanzella06}, \citet{vanzella08}, and \citet{vanzella09}. 
Most of these sources are Lyman-break galaxies photometrically selected using color criteria for spectroscopic follow-up. 
Our study takes a similar approach. 
Our MUSE spectroscopic sample is based upon optical+NIR HST detection and blind searches for emission lines in the spectral cubes. 
Then we verify them with photometric redshifts. 
As a result, it is unsurprising to find similar $\xi_{\textrm{ion,0}}$'s for comparable $M_{UV}$ in previous work. 

It is interesting to try quantify the dependence of $\xi_{\textrm{ion,0}}$ on M$_{UV}$.  To examine this, we assume $\log_{10} \xi_{ion,0}$ varies linearly with $M_{UV}$ and then fit to the present observations and also the spectroscopic sample of  \citet{bouwens16}.   The best-fit line we derive is $\xi_{\textrm{ion,0}} = 0.020 (\pm 0.031) ($M$_{UV}+20) + 25.372 (\pm 0.045)$.   While our best-fit favors a slight increase in $\xi_{\textrm{ion,0}}$ towards fainter M$_{UV}$'s, the trend is not statistically significant ($<$1$\sigma$).  For the fits, the binning we consider for the $z\approx5$ sample from \citet{bouwens16} has been to include at least 9 sources in each bin, with values of $\log_{10} \xi_{ion,0} = 25.43_{-0.06}^{+0.06}$ at $M_{UV}=-21.35$ and $\log_{10} \xi_{ion,0} = 25.63_{-0.18}^{+0.17}$ at $M_{UV}=-20.05$.

At low redshifts, \citet{shivaei18} measured $\xi_{\textrm{ion,0}}$ directly from the H$\alpha$ line fluxes of 676 galaxies at $1.4 < z < 2.6$ in the MOSDEF spectroscopic survey. 
By measuring the line flux of an additional Balmer line, H$\beta$, they are able to accurately correct for the dust attenuation of H$\alpha$. 
They measured a fairly constant $\log_{10} \xi_{\textrm{ion,0}}$ of 25.06 (25.34)\footnote{assuming a Calzetti (SMC) dust law} spanning absolute UV magnitudes from M$_{UV}$ $\approx$ $-$22.5 to M$_{UV}$ $\approx$ $-$19.5. 
\citet{matthee17} calculated $\xi_{\textrm{ion,0}}$ for 588 H$\alpha$ emitters and 160 Ly$\alpha$ emitters at $z=2.2$. 
Their sample was based on an H$\alpha$/Ly$\alpha$-selection using narrow-band images. 
They found an increasing trend in $\xi_{\textrm{ion,0}}$ with absolute UV magnitude from $\textrm{M}_{UV} = -23$ to $\textrm{M}_{UV} = -17$. 

\subsection{Tentative Detection of [S III] 9068.6\r{A} in the SEDs of $z\approx2.9-4.5$ Galaxies}

One interesting new finding in our analysis is the detection of [S III] 9068.6\r{A} in our stack results. 
It is detected in the stacks of three bins that contain the brightest sources, $-20.5<M_{UV,AB}<-19.5$ , $10 < \textrm{log}_{10} (\textrm{M}_{*}/\textrm{M}_{\odot}) < 9$, and $-1.5 < \beta < -2.0$ bins. 
Our analysis suggests the presence of [S III] 9068.6\r{A} with a rest-frame EW of $\approx$110\r{A}.  
Its presence can be seen as a `bump' at z$\approx$3.5 in the [3.6]-[4.5] color against redshift (see figure \ref{fig:colors_fit}). 

This line is mostly out of the spectral range of spectroscopic surveys at both low and intermediate redshifts which were conducted in the optical, but we can look at theoretical predictions for the line. 
\citet{anders03} predicts the flux in the [SIII] 9068.6\r{A} line to be 33\% as strong as H$\alpha$ for galaxies in the metallicity range 0.4-2.5 $Z_{\odot}$ and 18\% as strong as H$\alpha$ for galaxies with metallicities of 0.2 $Z_{\odot}$. 
Interestingly, in our brightest magnitude bin, the lines we infer at $\approx$9000$\,$\r{A} have a measured flux which is $\approx$11-16\% of the flux in the H$\alpha$ line, which is comparable with predictions of low metallicities.

\subsection{Implications for Reionization\label{sec:reion}}

In evaluating the capacity of star-forming galaxies to drive cosmic reionization, the total ionizing emissivity is typically calculated by multiplying three separate factors: the unattenuated UV luminosity density $\rho_{UV}/f_{\textrm{esc,UV}}$, the ionizing photon production efficiency $\xi_{\textrm{ion}}$, and the Lyman-continuum escape fraction $f_{\textrm{esc,LyC}}$. 
In the present analysis, we were able to place constraints on $\xi_{\textrm{ion,0}}$ for galaxies at a time shortly after reionization has completed. 
If we assume galaxies in the era of reionization ($6 < z < 9$) had similar efficiencies in producing ionizing photons as the ones we analyzed here ($z\sim4$-5), we can set limits on the escape fraction of ionizing photons in those galaxies. 

Assuming star-forming galaxies drive the reionization of the universe, \citet{bouwens15} have shown that the relative escape fraction $f_{\textrm{esc,rel}} \equiv f_{\textrm{esc,LyC}} / f_{\textrm{esc,UV}}$, and $\xi_{\textrm{ion}}$ must satisfy the following relation: 
\begin{equation}\label{eq:fesc_constraint}
f_{\textrm{esc,rel}} \hspace{0.1cm} \xi_{\textrm{ion}} \hspace{0.1cm} f_{\textrm{corr}}(M_{\textrm{lim}}) \hspace{0.1cm} (C/3)^{-0.3} = 10^{24.50\pm0.10} \hspace{0.1cm} \textrm{s$^{-1}$/(erg s$^{-1}$ Hz$^{-1}$), }
\end{equation}
where $M_{\textrm{lim}}$ is the assumed $UV$ luminosity cut off and $f_{\textrm{corr}}(M_{\textrm{lim}})$ is a correction factor for $\rho_{UV}(z=8)$ integrated to different values of $M_{\textrm{lim}}$.
\citet{bouwens15} found that $\log_{10}(f_{\textrm{corr}}(M_{\textrm{lim}}))$ could be approximated as $0.02+0.078(M_{\textrm{lim}}+13)-0.0088(M_{\textrm{lim}}+13)^{2}$. 
Note that $f_{\textrm{corr}}(M_{\textrm{lim}})$ is close to unity when $M_{\textrm{lim}}$ = $-$13, which is a typical limiting magnitude chosen by many studies \citep[e.g.,][]{robertson15,bouwens15}.  
The clumping factor, $C = \langle n^{2}_{H} \rangle / \langle n_{H} \rangle^{2}$, is commonly chosen to be 3, a value motivated by simulations \citep[e.g.,][]{pawlik09}.

As indicated by Equation \ref{eq:fesc_constraint}, we have a collective constraint on the product of the efficiency with which star-forming galaxies produce ionizing photons ($\xi_{\textrm{ion}}$) and the implied relative escape fraction ($f_{\textrm{esc,rel}}$). 
The product of these two factors cannot be greater than indicated by Equation \ref{eq:fesc_constraint} or the cosmic reionization would have been completed sooner than observed, i.e., at $z\approx6$.

\begin{figure}
  \centering
  \resizebox{\hsize}{!}
  {\includegraphics{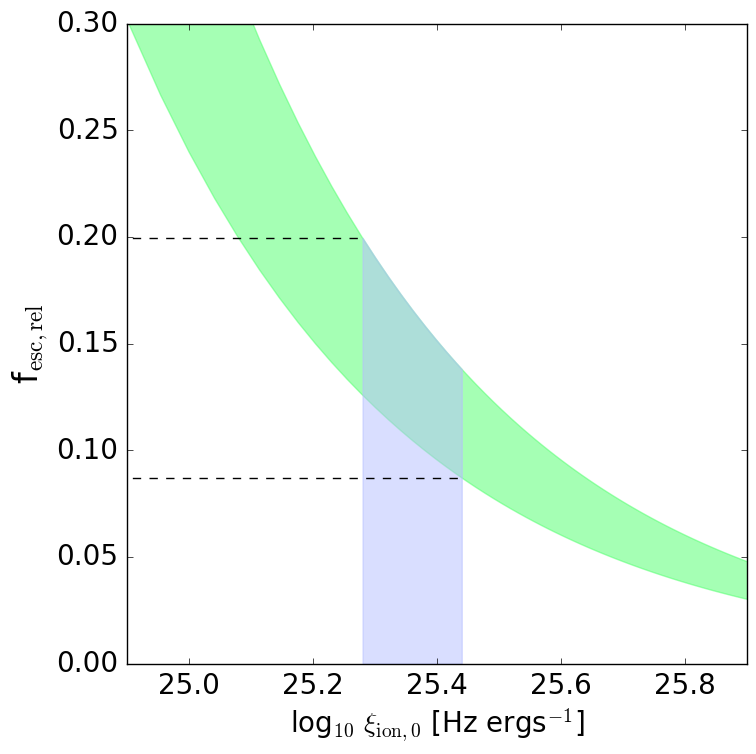}}
\caption{Relative escape fraction $f_{\textrm{esc,rel}} \equiv f_{\textrm{esc,LyC}} / f_{\textrm{esc,UV}}$ as a function of $\xi_{\textrm{ion,0}}$ is plotted in green, assuming star-forming galaxies solely drove reionization.  
Since the dust attenuation estimated for (non-ionizing) UV wavelengths is negligible ($f_{\textrm{esc,UV}}=1$), $f_{\textrm{esc,rel}}$ equals $f_{\textrm{esc,LyC}}$. 
We also approximated that, for small $f_{\textrm{esc,LyC}}$, $\xi_{\textrm{ion}} \approx \xi_{\textrm{ion,0}}$. 
The green region was derived in the analysis of \citet{bouwens15} by making use of important constraints on reionization (see Table 1 of \cite{bouwens15} for a complete list).  
The blue region shows the range of $\xi_{\textrm{ion,0}}$ inferred in the present study for faint galaxies, i.e. $\xi_{\textrm{ion,0}}$ = 25.36$\pm$0.08. 
The corresponding constraints we can place on $f_{\textrm{esc,rel}}$ of $\approx$8-20\% are indicated by the dashed lines.  
See \S\ref{sec:reion}.  \label{fig:fesc_vs_xi}}
\end{figure}

The average inferred $\log_{10} \xi_{\textrm{ion,0}}$ (in $\textrm{M}_{UV}$ bins) is 25.36$\pm$0.08. 
Given that faint galaxies provide the dominant contribution to the overall $UV$ luminosity density, we treat all galaxies as having the same $\xi_{\textrm{ion,0}}$ value as the fainter sources we are studying here. 
If we take the inferred value of $\xi_{\textrm{ion,0}}$ as typical, and approximate $\xi_{\textrm{ion}} = \xi_{\textrm{ion,0}} / (1-f_{\textrm{esc,LyC}}) \approx \xi_{\textrm{ion,0}}$ for small $f_{\textrm{esc,LyC}}$, we determine that the relative escape fraction cannot be larger than $\approx$8-20\% using Equation \ref{eq:fesc_constraint}. 
This relation and the constraint on the relative escape fraction are visually presented in Figure \ref{fig:fesc_vs_xi}. 

The average UV-continuum slope of sources within $-20.5 < M_{UV} < -17.5$ at 3.856 < $z$ < 5.329 is $\beta = -2.26$. 
This means the correction for dust extinction required is negligible ($f_{\textrm{esc,UV}} = 1$) according to the SMC dust law (equation \ref{eq:smc}). 
In this case, the relative escape fraction, $f_{\textrm{esc,rel}}$, equals the Lyman-continuum escape fraction, $f_{\textrm{esc,LyC}}$. 
Our constraints on $f_{\textrm{esc,rel}}$ (or $f_{\textrm{esc,LyC}}$) is consistent with the new Lyman-continuum escape fraction results from \citet{Steidel18}, which imply $f_{\textrm{esc,LyC}}$ values of $\approx0.09$ for sub-L* and brighter galaxies.

\section{Summary}

In this paper, we measured the EWs of H$\alpha$ and the Lyman-continuum photon production efficiency $\xi_{\textrm{ion,0}}$ for galaxies fainter than 0.2 $L^*$ in the redshift interval z$\sim$3-5.  
Because H$\alpha$ is a recombination line, its EW provides a useful measurement of $\xi_{\textrm{ion,0}}$, the intrinsic rate at which ionizing photons are produced per $UV$-continuum photon. 
Since faint galaxies likely dominated the ionizing photon budget, measurements of $\xi_{\textrm{ion,0}}$ for faint galaxies allow us to better understand the role of star-forming galaxies played in cosmic reionization. 

We are able to extend $\xi_{\textrm{ion,0}}$ measurements to uniquely low luminosities thanks to many spetroscopic measurements for faint sources from the MUSE GTO program and the 200-hour Spitzer/IRAC data now available from GREATS + other Spitzer/IRAC programs.  This combined data set constitutes the deepest available Spitzer/IRAC imaging over any part of the sky. 
To measure accurate IRAC photometry, we use the deep HST images as priors for modeling the surface brightness profiles of sources in IRAC images using the code MOPHONGO. 

The large number of spectroscopic redshifts we have available from the MUSE-Deep and MUSE-Wide programs allow us to segregate sources into a few distinct redshift intervals where specific strong emission lines fall within distinct Spitzer/IRAC filters (see Figure \ref{fig:halpha} and Table \ref{table:z_ranges}). 
From the redshift dependence of the [3.6]$-$[4.5] colors we can estimate the H$\alpha$ EW, and thus infer $\xi_{\textrm{ion,0}}$. 

Sources are subdivided into different subsamples according to their physical properties (M$_{UV}$, M$_{*}$, and $\beta$), and are stacked (see Equations \ref{eq:weight_components} and \ref{eq:weight_scheme}). 
To account for the potential contamination of [S III] 9068.6 \r{A}, we consider galaxies at lower redshifts that give us leverage on the [S III] equivalent width in the same manner as z$\approx$4-5 galaxies give on H$\alpha$. 
We fit the observed [3.6]-[4.5] colors across z$\approx$2.9 to z$\approx$5.3 to constrain, simultaneously, the H$\alpha$ EW, the [S III] EW, and the rest-frame optical continuum slope. 
We measure a rest-frame H$\alpha$ EW of 403-2818 \r{A} and a rest-frame [S III] EW of $\approx$110 \r{A}. 

From our inferred H$\alpha$ EWs, we estimate an average $\log_{10} \xi_{\textrm{ion,0}}$ of 25.36$\pm$0.08 for sources between $-20.5 < \textrm{M}_{UV} < -17.5$. 
As such, we have been able to estimate $\xi_{\textrm{ion,0}}$ for UV luminosities 1.5 mag fainter than was possible in \citet{bouwens16}, and 2.5 mag fainter than \citet{harikane18}.  
Combining our new results with those from previous studies that probe brighter magnitudes, we do not find any statistically significant ($<$1$\sigma$) dependence of $\xi_{\textrm{ion,0}}$ on the $UV$ luminosity of star-forming galaxies. 
The larger uncertainties at high redshifts, however, do not imply inconsistency with the trend found at lower redshifts \citep{matthee17}. 
If we take our derived $\xi_{\textrm{ion,0}}$'s as typical, they imply a relative escape fraction no higher than $\approx$8-20\% for faint galaxies.

Our $\xi_{\textrm{ion,0}}$ values are potentially biased high due to the selection of sources which show Ly$\alpha$ in emission in the MUSE data, potentially selecting those sources going through an active burst.  To control for this potential bias, we will investigate how the Lyman-continuum photon production efficiency depends on the EW of the Ly$\alpha$ emission line in a forthcoming paper and combine with the Ly$\alpha$ escape fraction.

\begin{acknowledgements}
We acknowledge useful discussions with Jorryt Matthee and Renske Smit, and also support from NASA grant NAG5-7697, NASA grant {\it HST}-GO-11563, and NWO vrij competitie grant 600.065.140.11N211. 
JR acknowledges support from the ERC starting grant 336736-CALENDS. 
\end{acknowledgements}

%
%

\begin{appendix}

\section{PSF matching across the various HST bands}\label{apdx:psf}

We ran \texttt{SExtractor} on the F850LP (`z')-band image with a set of parameters optimized for detecting stars that have intermediate brightness, as listed in table \ref{table:find_star_sex_param}. 
We select stars with \texttt{CLASS\_STAR} greater than or equal to 0.1, and AB magnitudes within 16.0 $\leq$ z$_{850,AB}$ $\leq$ 20.0. 
We choose the size of the PSFs to be 91$\times$91 pixels, or 5.46"$\times$5.46". 
Stars close to the edge (distance from the star's center to the edge less than 46 pixels) are excluded. 
We also require a minimum separation 1.365" (a quarter of the PSF size) between stars and pixels belonging to other sources. 
There are 43 stars in total within the MUSE-Deep and MUSE-Wide regions that meet these criteria.  For each of the HSF bandpasses, PSFs are created by weighting, shifting, normalizing, and median stacking cut-outs of the 43 stars.

\begin{figure*}
\centering
{\includegraphics[width=18cm]{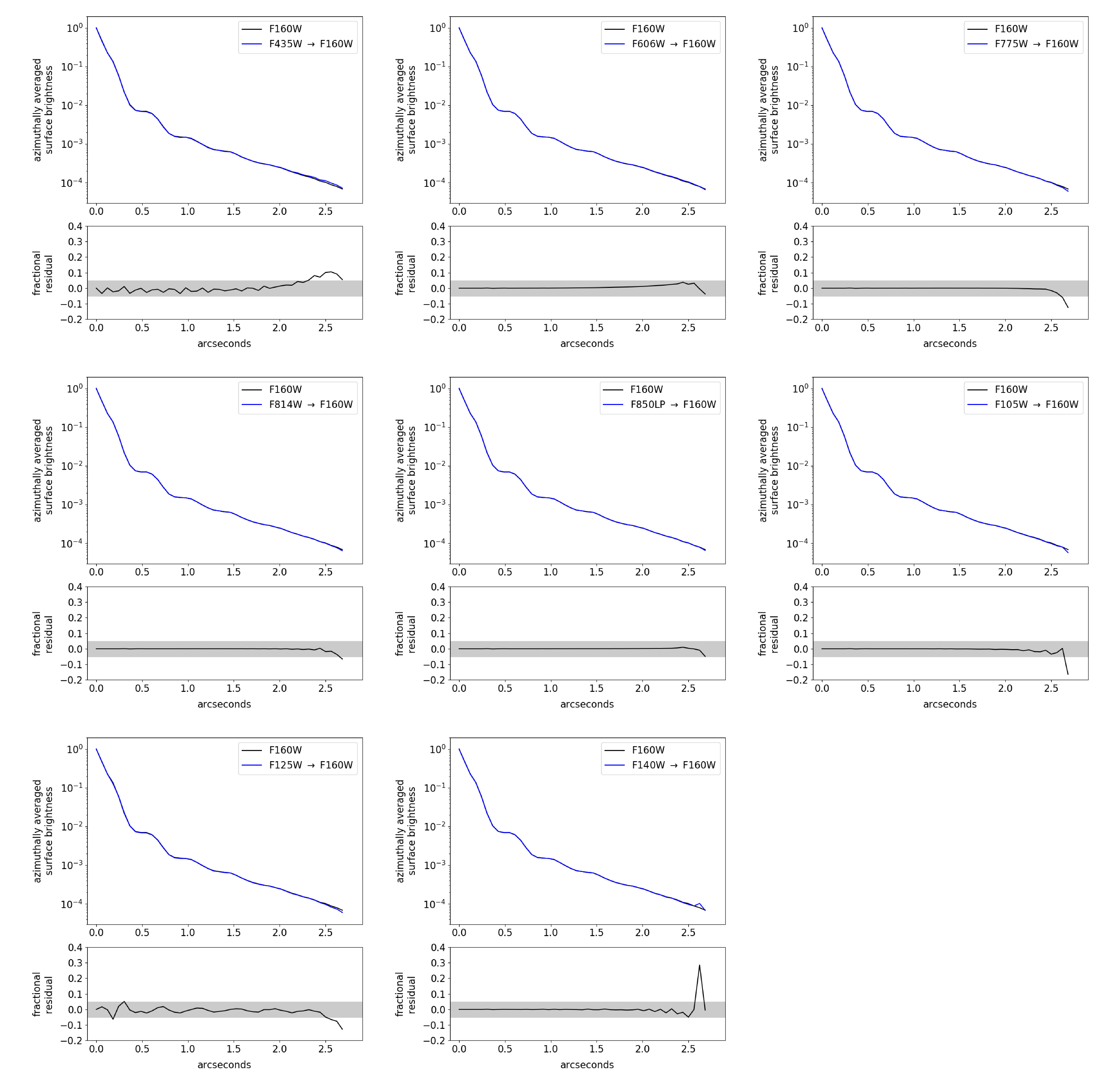}}
\caption{(\textit{upper sub-panels}) Comparison of the circularly-averaged radial profile in the F160W band (\textit{black curves}) with similar radial profiles of the PSFs in the other bands convolved with the corresponding best-fit kernels (\textit{blue curves)}.  (\textit{lower sub-panels}) The fractional residual between the profiles is shown.  The gray region denotes $\pm$5\%. }
\label{fig:psf}
\end{figure*}

Our PSF-matching kernels (from a given HST band to the F160W band) are constructed using the \texttt{create\_matching\_kernel} Python module from \texttt{photutils}.  Noise in the Fourier transforms is filtered with a `split cosine bell window', which has a parameter controlling the percentage of tapered values, and another controlling the fraction of the array size at which tapering begins. 
We search for the best kernel by fitting the reproduced radial profile of F160W PSF to the observed one. 
Figure \ref{fig:psf} shows the radial profiles of reproduced F160W PSFs. It is clear that the PSF-matched kernels are accurate to $\lesssim$1\% to a radius of 2.5". 

\section{SExtractor parameters}

In Table \ref{table:sex_param} and \ref{table:find_star_sex_param}, we present the \texttt{SExtractor} parameters we use for color measurements and the detection of stars.

\begin{table}[h!]
    \caption[Table caption text]{\texttt{SExtractor} parameters used for color measurements.}
	\begin{tabular}{ll}
		\label{table:sex_param}
	    parameter  &  value  \\
    	\hline
        \noalign{\smallskip}
	    \texttt{DETECT\_MINAREA}  &  5 \\
        \noalign{\smallskip}
	    \texttt{DETECT\_THRESH}  &  1.5 \\
        \noalign{\smallskip}
	    \texttt{ANALYSIS\_THRESH}  &  1.5 \\
        \noalign{\smallskip}
	    \texttt{DEBLEND\_NTHRESH}  &  32 \\
        \noalign{\smallskip}
	    \texttt{DEBLEND\_MINCONT}  &  0.001 \\
        \noalign{\smallskip}
        \texttt{CLEAN}  &  Y \\
        \noalign{\smallskip}
        \texttt{CLEAN\_PARAM}  &  1.0 \\
        \noalign{\smallskip}
        \texttt{BACK\_TYPE}  &  MANUAL \\
        \noalign{\smallskip}
        \texttt{BACK\_VALUE}  &  0.0 \\
        \hline
	\end{tabular}
\end{table}

\begin{table}[h!]
    \caption[Table caption text]{\texttt{SExtractor} parameters used for detecting stars.}
	\begin{tabular}{ll}
		\label{table:find_star_sex_param}
	    parameter  &  value  \\
    	\hline
        \noalign{\smallskip}
	    \texttt{DETECT\_MINAREA}  &  5 \\
        \noalign{\smallskip}
	    \texttt{DETECT\_THRESH}  &  3.5 \\
        \noalign{\smallskip}
	    \texttt{ANALYSIS\_THRESH}  &  3.5 \\
        \noalign{\smallskip}
	    \texttt{DEBLEND\_NTHRESH}  &  32 \\
        \noalign{\smallskip}
	    \texttt{DEBLEND\_MINCONT}  &  0.005 \\
        \noalign{\smallskip}
        \hline
	\end{tabular}
\end{table}

\section{Derivation of UV continuum slope}

\begin{table}[H]
    \caption[Table caption text]{HST filters used to estimate the UV continuum slope $\beta$ for sources in different redshift ranges}
	\begin{tabular}{cl}
		\label{table:beta_filters}
	    redshift  &  HST filters  \\
    	\hline
        \noalign{\smallskip}
	    2.514 < z < 2.65  &  F606W, F775W, F814W \\
        \noalign{\smallskip}
	    2.65 < z < 2.9  &  F606W, F775W, F814W, F850LP \\
        \noalign{\smallskip}
	    2.9 < z < 3.3  &  F606W, F775W, F814W, F850LP, F105W \\
        \noalign{\smallskip}
	    3.3 < z < 3.8  &  F775W, F814W, F850LP, F105W \\
        \noalign{\smallskip}
	    3.8 < z < 4.2  &  F775W, F814W, F850LP, F105W, F125W \\
        \noalign{\smallskip}
	    4.2 < z < 5.329  &  F105W, F125W, F140W \\
        \noalign{\smallskip}
        \hline
	\end{tabular}
\end{table}

\section{Distribution of the weights utilized in our stack results}

\begin{figure}
\centering
{\includegraphics[width=7cm]{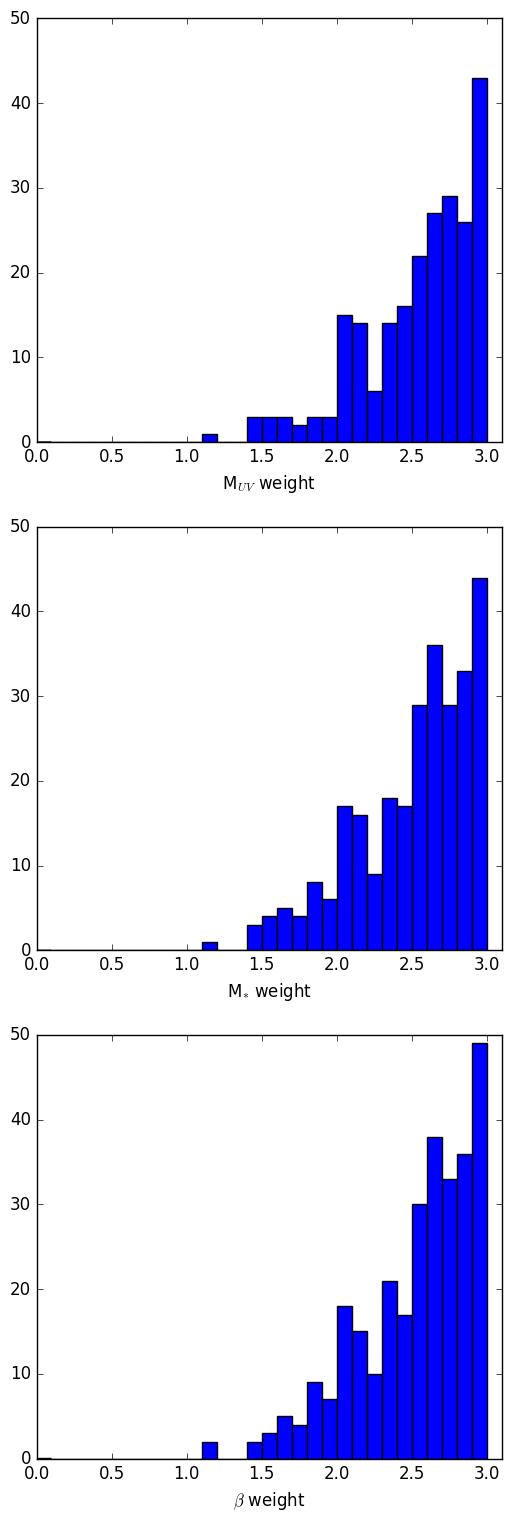}}
\caption{Distribution of total weights used for individual sources in the $M_{UV}$-, $M_{*}$-, and $\beta$-binned stacks (see Section \ref{sec:stack}).}
\label{fig:weights}
\end{figure}

In coadding the `cleaned' IRAC images from many sources to create deep stacks, we weight images in the stacks according to the quality of the stacks, contamination from the neighbors, and estimated flux uncertainties in the stack (see section \ref{sec:stack}).  To illustrate the range of weights (i.e., $w = w_{\chi^2} + w_{err} + w_{cont}$) we use in  stacking sources as a function of $M_{UV}$, $M_{*}$, and $\beta$, we show histograms of the weight distributions in Figure~\ref{fig:weights}.

\section{Illustration of the best-fit spectra implied for different IRAC stacks in our analysis}

\begin{figure*}
\centering
\includegraphics[width=180mm]{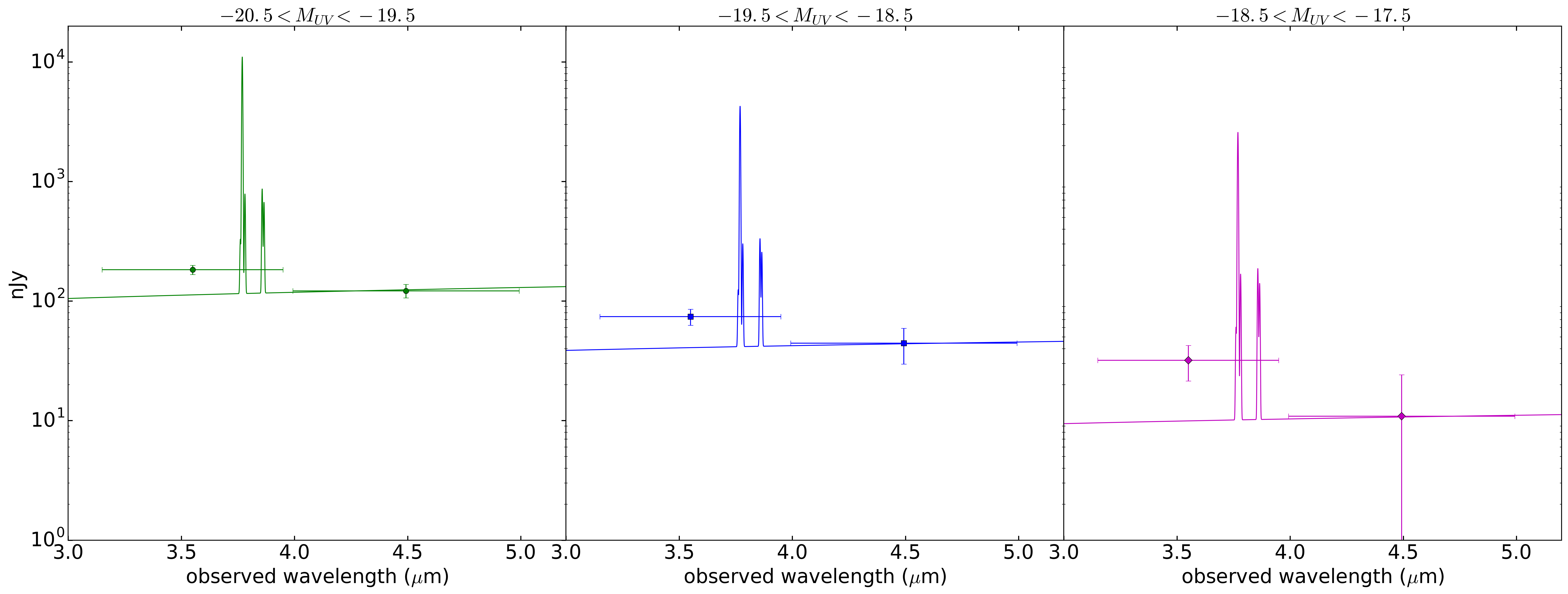}
\caption{An illustration of our stacked photometric measurements (\textit{solid circles}) in the 3.6$\mu$m and 4.5$\mu$m bands for galaxies at $4.532<z<4.955$ in three different bins of $UV$ luminosity $M_{UV}$ $-20.5<M_{UV}<-19.5$ (\textit{left}), $-19.5<M_{UV}<-18.5$ (\textit{center}), and $-18.5<M_{UV}<-17.5$ (\textit{right}).  The width of the 3.6$\mu$m and 4.5$\mu$m bands are shown with horizontal error bars, while the vertical error bars are $1\sigma$.  The lines show example model spectra using the line ratios assumed in our analysis (\S 4.2). }
\label{fig:muv_stack_spectra}
\end{figure*}

In deriving the implied H$\alpha$ EWs for sources, we make use of deep stacks of the Spitzer/IRAC observations from a large number of sources.  To provide our readers of how the SED stacks for individual sources look, we have included Figure~\ref{fig:muv_stack_spectra} showing the IRAC stack results for galaxies in the redshift range $4.532<z<4.955$ for three different bins in $UV$ luminosity $-20.5<M_{UV}<-19.5$, $-19.5<M_{UV}<-18.5$, and $-18.5<M_{UV}<-17.5$.

\section{Consistency check with \cite{smit16}}

As a test on the consistency of the H$\alpha$ luminosities inferred in this study with other work, we consider the \cite{smit16} study which derived H$\alpha$ luminosities for many sources over the GOODS North and South fields and conduct a comparison.  We compute the H$\alpha$ luminosity from the color difference:
\begin{align*}
\Delta c &= ([3.6]-[4.5]) - ([3.6]-[4.5])_{cont} \\*
&= -2.5 \textrm{ log}_{10} \frac{f_{3.6}+f_{H\alpha}}{f_{4.5}} + 2.5 \textrm{ log}_{10} (\frac{f_{3.6}}{f_{4.5}})_{cont} \\
&= -2.5 \textrm{ log}_{10} (1+\frac{f_{H\alpha}}{f_{3.6}}) \\
f_{H\alpha} &= (10^{\Delta c/-2.5} -1) \times f_{3.6} \\
&= (10^{\Delta c/-2.5} -1) \times f_{4.5} \times 10^{([3.6]-[4.5])_{cont}/-2.5}
\textrm{ ,}
\end{align*}
where ([3.6]-[4.5])$_{cont}$ = 0.04 is the average continuum color derived from bright ($-20.8 < M_{UV} < -19.4$) sources at 2.514 < z < 3.829, and assume f$_{3.6}$/f$_{4.5}$ equals (f$_{3.6}$/f$_{4.5}$)$_{cont}$. 
\cite{smit16} compute the H$\alpha$ luminosities from the difference between the observed f$_{3.6}$ and the f$_{3.6}$ expected from a best-fit SED without emission lines. 
Table \ref{table:consistency_check} compares our derived H$\alpha$ luminosities with those of the overlapping sources in \cite{smit16}. 
We find good agreement with \cite{smit16} within 1$\sigma$ for most sources. 

\begin{table}[!h]
    \caption[Table caption text]{Comparison of the H$\alpha$ luminosities derived in this work with \cite{smit16}. }
	\begin{tabular}{lll}
		\label{table:consistency_check}
         & \multicolumn{2}{c}{H$\alpha$ Luminosity (erg/s)} \\
	    ID  &  This Work  &  \cite{smit16}  \\
    	\hline
        \noalign{\smallskip}
	    deep\_1264  &  6.2$\pm$0.8$\times$10$^{42}$  & 4.4$\pm$1.0$\times$10$^{42}$  \\
        \noalign{\smallskip}
        wide\_125040110  &  6.4$\pm$1.0$\times$10$^{42}$  & 5.3$\pm$1.0$\times$10$^{42}$  \\
        \noalign{\smallskip}
        wide\_104032086  &  2.1$\pm$0.8$\times$10$^{42}$  & 1.6$\pm$ 1.2$\times$10$^{42}$  \\
        \noalign{\smallskip}
        deep\_1343  &  0.9$\pm$0.4$\times$10$^{42}$  & 1.2$\pm$0.8$\times$10$^{42}$  \\
        \noalign{\smallskip}
        deep\_68  &  2.3$\pm$0.7$\times$10$^{42}$  & 1.9$\pm$0.9 $\times$ 10$^{42}$  \\
        \noalign{\smallskip}
        \hline
	\end{tabular}
\end{table}

\end{appendix}

\end{document}